\documentclass[aps,showpacs,nofootinbib,superscriptaddress]{revtex4-1}
\usepackage{epsf}
\usepackage{psfrag}
\usepackage{graphicx}
\usepackage{amsmath}
\def\slashchar#1{\setbox0=\hbox{$#1$}
   \dimen0=\wd0 \setbox1=\hbox{/} \dimen1=\wd1
   \ifdim\dimen0>\dimen1 \rlap{\hbox to \dimen0{\hfil/\hfil}} #1
   \else  \rlap{\hbox to \dimen1{\hfil$#1$\hfil}} / \fi}

\newcommand{\be}{\begin{equation}}
\newcommand{\ee}{\end{equation}}
\newcommand{\bea}{\begin{eqnarray}}
\newcommand{\eea}{\end{eqnarray}}

\def\g#1{\gamma_{#1}}
\makeatletter
\def\slashchar#1{{\mathpalette\c@ncel{#1}}} 
\makeatother

\begin{document}

\title{ Exclusive $c\to s,d$ semileptonic decays of ground-state
spin-1/2 and spin-3/2 doubly heavy $cb$ baryons }

\author{ C.Albertus} \affiliation{Departamento de F\'\i sica Fundamental e
IUFFyM,\\ Universidad de Salamanca, E-37008 Salamanca, Spain} \author{
E. Hern\'andez} \affiliation{Departamento de F\'\i sica Fundamental e
IUFFyM,\\ Universidad de Salamanca, E-37008 Salamanca, Spain}
\author{J.~Nieves} \affiliation{Instituto de F\'\i sica Corpuscular
(IFIC), Centro Mixto CSIC-Universidad de Valencia, Institutos de
Investigaci\'on de Paterna, Apartado 22085, E-46071 Valencia, Spain}

\pacs{12.39.Jh, 13.30.Ce, 14.20.Mr}

\begin{abstract}
  We evaluate exclusive semileptonic decays of ground-state spin-1/2
  and spin-3/2 doubly heavy $cb$ baryons driven by a $c\to s,d$ transition at
  the quark level.  We check our results for the form factors against
   heavy quark spin symmetry constraints  obtained in the
  limit of  very large heavy quark masses  and near zero recoil.
  Based on those constraints we make  model independent,
though approximate,   predictions for ratios of decay widths.
\end{abstract}

\maketitle
%
%
%
%
%
%
\section{Introduction}
In this work we make a systematic analysis of exclusive semileptonic
$c\to s,d$ decays of doubly heavy ground state $cb$ baryons. Previous
studies are very limited and, to our knowledge, they only include the
work in Ref.~\cite{sanchis95} where the $\Xi'_{cb}\to\Xi_b$ decay was
analyzed using heavy quark spin symmetry, the relativistic three quark
model calculation of the $\widehat\Xi_{cb}\to\Xi'_b$ decay in
Ref.~\cite{Faessler:2001mr}, and the combined branching ratio for the
$(\Xi_{cb}\to\Xi_b)+(\Xi_{cb}\to\Xi'_b)+(\Xi_{cb}\to\Xi^*_b)$ decay
evaluated in Ref.~\cite{Kiselev:2001fw} in the framework of the
potential approach and QCD sum rules\footnote{In the case of the
  $\widehat\Xi_{cb}$ baryon, the spin of the $cn$ ($n=u,d$) pair is
  well defined and it is coupled to one. For the $\Xi_{cb}$ and
  $\Xi'_{cb}$ states, it is however the spin of the two heavy quarks
  ($cb$) the one which is well defined, $1$ and $0$, respectively (see
  Table~\ref{tab:baryons}). The different spin configurations are
  discussed in detail in Sect.~\ref{sect:mix}.}. Since the modulus of
the Cabbibo-Kobayashi-Maskawa (CKM) matrix elements
$|V_{cs}|,\,|V_{cd}|$ are much larger than $|V_{cb}|$, one would
 expect the decay widths for $c\to s,d$ semileptonic decay of
$cb$ baryons to be much larger than the corresponding $b\to c$ driven
decays which have been more extensively studied in the
literature~\cite{sanchis95,Ebert:2004ck,Hernandez:2007qv,pervin2,
  Faessler:2009xn,Albertus:2009ww}.  However, this is corrected by a
smaller available phase space, and the decay widths for $c\to s$
transitions turn out to be larger but of the same order of magnitude
as the $b\to c$ decay widths, while widths for $c\to d$ transitions
are much smaller. In any case, the analysis of the $c\to s,d$ decays of
$cb$ baryons could give relevant information on heavy quark physics
complementary to the one obtained from the study of the $b\to c$
decays.

Similar to what happens in atomic physics, in hadrons with a single
heavy quark the dynamics of the light degrees of freedom becomes
independent of the heavy quark flavor and spin when the mass of the
heavy quark is much larger than $\Lambda_{QCD}$ and the masses and
momenta of the light quarks. This is the essence of heavy quark
symmetry (HQS)~\cite{hqs1,hqs2,hqs3,hqs4}. HQS guaranties that in a
heavy baryon the light degrees of freedom quantum numbers are well
defined. Then, up to corrections in the inverse of the heavy quark
mass, one can take the spin of the two light quarks to be well
defined.  The two light quarks couple to a state with spin $S=$0 or 1
and then couple with the $b$ quark to total spin 1/2 or 3/2.  This is
the classification scheme followed for the $b$ heavy baryons in
Table~\ref{tab:baryons}.  However, HQS can not be applied to hadrons
containing two heavy quarks. There, the kinetic energy term needed to
regulate infrared divergences breaks the heavy quark flavor symmetry,
but not the spin symmetry for each heavy quark
flavor~\cite{thacker91}. This is known as heavy quark spin symmetry
(HQSS). According to HQSS~\cite{Jenkins:1992nb}, for large heavy quark
masses one can select the heavy quark subsystem of a doubly heavy
baryon to have a well defined total spin.  Again this is the
classification scheme followed for $cb$ states shown in
Table~\ref{tab:baryons}. There, the $c$ and $b$ quark couple to a state
with spin $S=$0 or 1 and then couple with the light quark to total spin
1/2 or 3/2.  Being the heavy quark masses finite, one has that for
spin-1/2 baryons the hyperfine interaction can admix both $S=$0 and $S$=1
components into the wave function of physical states. As shown in
Sec.~\ref{sect:mix} this is very relevant for spin-1/2 $cb$
baryons. In principle one should also expect some degree of mixing for
the $\Xi_b$ and $\Xi'_b$ states. However, in this latter case the
hyperfine matrix elements responsible for mixing are proportional to
the inverse of the $b$ quark mass and mixing effects are thus
suppressed.

In Table~\ref{tab:baryons} we present the baryons involved in the
present study.  As mention the $\Xi_{cb},\Xi'_{cb}$ and
$\Omega_{cb},\Omega'_{cb}$ are not the physical states that will be
discussed in the following. The quark model masses in
Table~\ref{tab:baryons} have been taken from our previous works in
Refs.~\cite{Albertus:2003sx,Albertus:2009ww}, where they were obtained
using the AL1 potential of
Refs.~\cite{semay94,silvestre96}. Experimental masses are the ones
given by the particle data group (PDG) in Ref.~\cite{pdg10} and in the
table we quote the average over the different charge states. The
agreement with our results is better than 1\%. For the actual
calculation of the decay widths we shall use experimental masses taken
from Ref.~\cite{pdg10} whenever possible.  For the neutral
$\Sigma^{*\,0}_b$ state we follow Ref.~\cite{hwang07} and take
$M_{\Sigma_b^{*\,0}}=\frac12(M_{\Sigma_b^{*\,+}}+M_{\Sigma_b^{*\,-}})
$. For the $\Sigma_b^{0}$ case, corrections to the analogous relation,
due to the electromagnetic interaction between the two light quarks in
the heavy baryon, have been evaluated in Ref.~\cite{Hwang:2008dj}
using heavy quark effective theory and in Ref.~\cite{Guo:2008ns} in
chiral perturbation theory to leading one-loop order. Based on the
known experimental data they get
$M_{\Sigma_b^{0}}=5810.5\pm2.2\,$MeV~\cite{Hwang:2008dj} and
$M_{\Sigma_b^{0}}=5810.3\pm1.9\,$MeV~\cite{Guo:2008ns}, their central
values being 1\,MeV lower than the value one would obtain from the
less accurate relation $M_{\Sigma_b^{\,0}
}=\frac12(M_{\Sigma_b^{\,+}}+M_{\Sigma_b^{\,-}})$. Here we shall use
the value $M_{\Sigma_b^{0}}=5810.5\,$MeV given in
Ref.~\cite{Hwang:2008dj}.  For the $\Xi'_b,\Xi^*_b,\Omega^*_b$ we take
our predictions in Ref.~\cite{Albertus:2003sx} which are in agreement
with lattice results by the UKQCD Collaboration~\cite{ukqcd96}. For
doubly heavy $cb$ baryons baryons there is no experimental information
on their masses and we shall use our own predictions in
Ref.~\cite{Albertus:2009ww}.

\begin{table}
\begin{tabular}{ccccccc}\hline\hline
Baryon &~~~~$J^P$~~~~&~~~~ $I$~~~~&~~~~$S^\pi$~~~~& 
Quark content &\multicolumn{2}{c}{Mass\ [MeV]}\\\cline{6-7}

       &       &         &   &      & Quark model   & Experiment                
\\       &       &         &   &      & 
\cite{Albertus:2003sx,Albertus:2009ww}   & \cite{pdg10} \\  
\hline
$\Xi_{cb}$ &$\frac12^+$& $\frac12$ &$1^+$&$cbn$&6928&--
\\
$\Xi'_{cb}$ &$\frac12^+$& $\frac12$ &$0^+$&$cbn$&6958&--
\\
$\Xi^*_{cb}$ &$\frac32^+$& $\frac12$ &$1^+$&$cbn$&6996&--
\\
$\Omega_{cb}$  &$\frac12^+$& 0 &$1^+$&$cbs$&7013&--\\
$\Omega'_{cb}$  &$\frac12^+$& 0 &$0^+$&$cbs$&7038&--\\
$\Omega^*_{cb}$  &$\frac32^+$& 0 &$1^+$&$cbs$&7075&--\\

\hline
$\Lambda_b$ &$\frac12^+$& 0 &$0^+$&$udb$&5643&$5620.2\pm1.6$
\\
$\Sigma_b$ &$\frac12^+$& 1 &$1^+$&$nnb$&5851&$5811.5\pm2.4$
\\
$\Sigma^*_b$ &$\frac32^+$& 1 &$1^+$&$nnb$&5882&$5832.7\pm3.1$
\\
$\Xi_b$  &$\frac12^+$&$\frac12$&$0^+$&$nsb$&5808&$5790.5\pm2.7$
\\
$\Xi'_b$  &$\frac12^+$&$\frac12$&$1^+$&$nsb$&5946&--
\\
$\Xi^*_b$ &$\frac32^+$&$\frac12$&$1^+$&$nsb$&5975&--
\\
$\Omega_b$  &$\frac12^+$& 0 &$1^+$&$ssc$&6033&$6071\pm40$
\\
$\Omega^*_b$  &$\frac32^+$& 0 &$1^+$&$ssc$&6063&--
\\\hline
\hline
\end{tabular}
\caption{Quantum numbers of baryons involved in this study. For the $cb$
baryons,
 states with a well defined spin for the heavy subsystem are shown.  $J^\pi$ and $I$ are the spin-parity and
 isospin of the baryon, while $S^\pi$ is the spin-parity of the two
 heavy or the two light quark subsystem. $n$ denotes a $u$ or $d$
 quark. Experimental masses are isospin averaged over the values reported by the
 PDG~\cite{pdg10}.}
\label{tab:baryons}
\end{table}
The paper is organized as follows: in Sec~\ref{sect:mix} we discuss
the physical spin-1/2 $cb$ baryons and the relevance of hyperfine
mixing for those states. In Sec.~\ref{sect:dwff} we give general
formulas needed to compute the semileptonic decay width, we present
the form factor decompositions that we use for the different
transitions and we present and discuss our predictions for the $c\to
s,d$ decay widths.  In Sec.~\ref{sect:hqss} we obtain HQSS constraints
for the form factors and make predictions for ratios of decay widths
based on those constraints. Finally in Sect.~\ref{sec:summa}, we
summarize the main results of this work. The paper contains also two
appendices. In appendix~\ref{app:nrbs} we present our nonrelativistic
baryon states, while in appendix~\ref{app:ffwme} we give details on how
we evaluate the transition matrix elements and form factors.

\section{Configuration mixing in $cb$ doubly heavy baryons}
\label{sect:mix}
Due to the finite value of the heavy quark masses, the hyperfine
interaction between the light quark and any of the heavy quarks can
admix both $S$=0 and 1 components into the wave function for total
spin-1/2 states. Thus, the actual physical spin-1/2 $cb$ baryons are
admixtures of the $\Xi_{cb},\,\Xi'_{cb}$
($\Omega_{cb},\,\Omega'_{cb}$) states listed in
Table~\ref{tab:baryons}.  The physical states, that we shall call
$\Xi_{cb}^{(1)},\,\Xi_{cb}^{(2)}$ and
$\Omega_{cb}^{(1)},\,\Omega_{cb}^{(2)}$, are given within the AL1
model by~\cite{Albertus:2009ww}\footnote{Note that, here we use the
  order $cb$, while in \cite{Albertus:2009ww}, we used $bc$. Thus our
  $\Xi'_{cb}$ and $\Omega'_{cb}$ states, where the heavy quark
  subsystem is coupled to zero,
  differ in one sign with those used in \cite{Albertus:2009ww}.}
\begin{eqnarray}
\Xi_{cb}^{(1)}&=&-0.902\ \Xi'_{cb}+0.431\ \Xi_{cb}\ \ ;\ M_{\Xi_{cb}^{(1)}}=6967\,{\rm MeV},\nonumber\\
\Xi_{cb}^{(2)}&=&\hspace{.25cm}0.431\  \Xi'_{cb}+0.902\ \Xi_{cb}\ \ ;\ M_{\Xi_{cb}^{(2)}}=6919\,{\rm MeV},
\label{eq:mixedxi}
\end{eqnarray}
\begin{eqnarray}
\Omega_{cb}^{(1)}&=&-0.899\  \Omega'_{cb}+0.437\ \Omega_{cb}\ \ ;\ M_{\Omega_{cb}^{(1)}}=7046\,{\rm MeV},\nonumber\\
\Omega_{cb}^{(2)}&=&\hspace{.25cm}0.437\  \Omega'_{cb}+0.899\ \Omega_{cb}\ \ ;
\ M_{\Omega_{cb}^{(2)}}=7005\,{\rm MeV},
\label{eq:mixedomega}
\end{eqnarray} 
Comparing the masses of the physical states with the mass values
quoted in Table~\ref{tab:baryons}, one sees that masses are not very
sensitive to hyperfine mixing. On the other hand, it was pointed out
by Roberts and Pervin~\cite{pervin1} that hyperfine mixing could
greatly affect the decay widths of doubly heavy $cb$ baryons.  This
assertion was checked in Ref.~\cite{pervin2} where Roberts and Pervin
found that hyperfine mixing in the $cb$ states has a tremendous impact
on doubly heavy baryon $b\to c$ semileptonic decay widths. These
results were qualitatively confirmed by our own calculation in
Ref.~\cite{Albertus:2009ww}. We further investigated the role of
hyperfine mixing in electromagnetic transitions~\cite{Albertus:2010hi}
finding again large corrections to the decay widths. A similar study
was conducted by Branz et al. in Ref.~\cite{Branz:2010pq}.  We expect
configuration mixing should also play an important role for $c\to s,d$
semileptonic decay of $cb$ baryons.

One way of minimizing the hyperfine mixing for $cb$ baryons is to use from
the start baryon states in which the $c$ quark and the light $q$ quark
couple to a state of well defined spin $S_{cq}=0$ or 1. Then the $b$
quark couples to that state to make the baryon with total spin 1/2. We
denote those states as $\widehat\Xi_{cb},\,\widehat\Omega_{cb}$ for
$S_{cq}=1$, and $\widehat\Xi'_{cb},\,\widehat\Omega'_{cb}$ for
$S_{cq}=0$. The relation between the latter set of states and the ones
in Table~\ref{tab:baryons} is given by (here $B$ stands for $\Xi$ or
$\Omega$)
\begin{eqnarray}
\widehat B_{cb}&=&-\frac{\sqrt3}{2}B'_{cb}+\frac{1}{2}B_{cb},\nonumber\\
\widehat B'_{cb}&=&\frac{1}{2}B'_{cb}+\frac{\sqrt3}{2}B_{cb}.
\label{eq:qchqss}
\end{eqnarray}
Hyperfine mixing for the $\widehat B_{cb},\,\widehat B'_{cb}$ states
is much less important since it is inversely proportional to the $b$
quark mass~\cite{Albertus:2009ww}. Physical spin-1/2 $cb$ baryons
states should then be very close to the $\widehat B_{cb},\,\widehat
B'_{cb}$ states and this is indeed the case.  If we write \bea
\left(\begin{array}{c}B^{(1)}_{cb}\vspace{.1cm}\\B^{(2)}_{cb}\end{array}\right)=
\left(\begin{array}{cc}\cos\theta&\sin\theta\\-\sin\theta&\cos\theta\end{array}\right)
  \left(\begin{array}{c}\widehat B_{cb}\\\widehat
    B'_{cb}\end{array}\right) \eea we find
  $\theta_\Xi=-4.46^o,\ \theta_\Omega=-4.07^o$ for the AL1 interquark
  interaction~\cite{Albertus:2009ww}.

\section{Semileptonic decay widths}
\label{sect:dwff}
\subsection{General formulas}
The total decay width for semileptonic $c\to l$ transitions, with
$l=s,d$, is given by 
\bea \Gamma&=&|V_{cl}|^2
\frac{G_F^{\,2}}{8\pi^4}\frac{M'^2}{M} \int\sqrt{\omega^2-1}\, {\cal
L}^{\alpha\beta}(q) {\cal H}_{\alpha\beta}(P,P')\,d\omega,
\eea 
where $|V_{cl}|$ is the modulus of the corresponding CKM matrix
element for a semileptonic $c\to l$ decay ($|V_{cs}|=0.97345$ and
$|V_{cd}|=0.2252$~\cite{pdg10}), $G_F= 1.16637(1)\times
10^{-11}$\,MeV$^{-2}$~\cite{pdg10} is the Fermi decay constant, $P,M$
($P',M'$) are the four-momentum and mass of the initial (final)
baryon, $q=P-P'$ and $\omega$ is the product of the initial and final
baryon four-velocities $\omega=v\cdot v'=\frac{P}{M} \cdot
\frac{P'}{M'}=\frac{M^2+M'^2-q^2}{2MM'}$. In the decay, $\omega$
ranges from $\omega=1$, corresponding to zero recoil of the final
baryon, to a maximum value that, neglecting the neutrino mass, is
given by $\omega=\omega_{\rm max}= \frac{M^2 + M'^2-m^2}{2MM'}$, which
depends on the transition and where $m$ is the final charged lepton
mass. Finally ${\cal L}^{\alpha\beta}(q)$ is the leptonic tensor after
integrating in the lepton momenta. It can be cast as
\bea
{\cal L}^{\alpha\beta}(q)=A(q^2)\,g^{\alpha\beta}+
B(q^2)\,\frac{q^\alpha q^\beta}{q^2},
\label{eq:lt}
\eea
where  explicit expressions for the scalar functions $A(q^2)$ and
$B(q^2)$ can be found in Eqs.~(3) and (4) of Ref.~\cite{Albertus:2011xz}.

The hadron tensor ${\cal H}_{\alpha\beta}(P,P')$ is given by
\begin{eqnarray}
{\cal H}^{\alpha\beta}(P,P') &=& \frac{1}{2J+1} \sum_{r,r'}  
 \big\langle B', r'\
\vec{P}^{\,\prime}\big| J_{cl}^\alpha(0)\big| B, r\ \vec{P}   \big\rangle 
\ \big\langle B', r'\ 
\vec{P}^{\,\prime}\big|J_{cl}^\beta(0) \big|  B, r\ \vec{P} \big\rangle^*,
\label{eq:wmunu}
\end{eqnarray}
with $J$ the initial baryon spin, $\big|B, r\ \vec P\big\rangle\,
\left(\big|B', r'\ \vec{P}\,'\big\rangle\right)$ the initial (final)
baryon state with three-momentum $\vec P$ ($\vec{P}\,'$) and spin
third component $r$ ($r'$) in its center of mass
frame\footnote{Baryonic states are normalized such that \bea
  \big\langle B, r'\ \vec{P}'\, |\,B, r \ \vec{P} \big\rangle =
  2E\,(2\pi)^3 \,\delta_{rr'}\, \delta^3 (\vec{P}-\vec{P}^{\,\prime}),
  \eea with $E$ the baryon energy for three-momentum $\vec P$.}. Our
states are constructed in appendix~\ref{app:nrbs}. Finally,
$J_{cl}^\mu(0)=\bar\Psi_{l}(0)\gamma^\mu(1-\gamma_5)\Psi_c(0)$ is the
$c\to l$ charged weak current.
\subsection{Form factors for  $1/2\to 1/2$,  $1/2\to 3/2$ and $3/2\to 1/2$ transitions}
For the actual calculation of the decay width we parametrize the hadronic 
matrix elements  in terms of form factors, which are functions of $\omega$
 or equivalently of $q^2$. The different form factor
decomposition that we use are given in the following.
\begin{enumerate} 
\item $\ 1/2 \to 1/2$ transitions.\\ 
Here we take the commonly used decomposition in terms of  three vector
  $F_1,\,F_2,\,F_3$ and three axial $G_1,\,G_2,\,G_3$ form factors
\begin{eqnarray}
\label{eq:1212}
\big\langle B'(1/2), r'\ \vec{P}^{\,\prime}\left|\,
J_{cl}^\mu(0) \right| B(1/2), r\ \vec{P}
\big\rangle& =& {\bar u}^{B'}_{r'}(\vec{P}^{\,\prime})\Big\{
\gamma^\mu\left[F_1(\omega)-\gamma_5 G_1(\omega)\right]+ v^\mu\left[F_2(\omega)-\gamma_5
G_2(\omega)\right]\nonumber\\
&&\hspace{1.5cm}+v'^\mu\left[F_3(\omega)-\gamma_5 G_3(\omega)
\right]\Big\}u^{B}_r(\vec{P}\,).\label{eq:def_ff}
\end{eqnarray}
 The $u_{r}$ are Dirac spinors normalized as $({ u}_{r'})^\dagger u_r
 = 2E\,\delta_{r r'}$.

\item $\ 1/2 \to 3/2$ transitions.\\
In this case we follow Llewellyn
 Smith~\cite{Llewellyn Smith:1971zm} to write
\begin{eqnarray}
\label{eq:1232}
&&\hspace{-1cm}\big\langle B'(3/2),r'\vec P'\,|\,\overline 
\Psi_l(0)\gamma^\mu(1-\gamma_5)\Psi_c(0)\,|\,B(1/2),r\,
\vec P\,\big\rangle=
~\bar{u}^{B'}_{\lambda\,r'}(\vec{P}\,')\,\Gamma^{\lambda\mu}(P,P')\,
u^{B}_r(\vec{P}\,),
\nonumber\\
\Gamma^{\lambda\mu}(P,P')=&&
\left[\frac{C_3^V}{M}(g^{\lambda\,\mu}q
\hspace{-.15cm}/\,
-q^\lambda\gamma^\mu)+\frac{C_4^V}{M^2}(g^{\lambda\,\mu}q\cdot
P'-q^\lambda
P'^\mu)+\frac{C_5^V}{M^2}(g^{\lambda\,\mu}q\cdot P-q^\lambda
P^\mu)+C_6^Vg^{\lambda\,\mu}\right]\gamma_5\nonumber\\
&&+\left[\frac{C_3^A}{M}(g^{\lambda\,\mu}q
\hspace{-.15cm}/\,
-q^\lambda\gamma^\mu)+\frac{C_4^A}{M^2}(g^{\lambda\,\mu}q\cdot P'-q^\lambda
P'^\mu)+{C_5^A}g^{\lambda\,\mu}+\frac{C_6^A}{M^2}
q^\lambda q^\mu\right].
\end{eqnarray}
Here $u^{B'}_{\lambda\,r'}$ is the Rarita-Schwinger spinor of the final spin
3/2 baryon normalized such that $(u_{\lambda\,r'}^{B'})^{\dagger}
u^{B'\,\lambda}_r = -2E'\,\delta_{rr'}$, and we have four vector
($C^V_{3,4,5,6}(\omega)$) and four axial ($C^A_{3,4,5,6}(\omega)$) form
factors. Within our model we shall have that 
$C^V_{5}(\omega)=C^V_{6}(\omega)=C^A_{3}(\omega)=0$.
 
\item $3/2\to 1/2$ transitions.\\
Similar to the case before we use
\begin{eqnarray}
&&\hspace{-1.5cm}\left\langle B'(1/2), r'\ \vec{P}^{\,\prime}\left|\,\overline \Psi_{l'}(0)\gamma^\mu(1-\gamma_5)
\Psi_c(0)
 \right| B(3/2), r\ \vec{P}
\right\rangle =
(\bar{u}^{B}_{\lambda\,r}(\vec{P}\,)\tilde\Gamma^{\lambda\,\mu}(P',P)
u^{B'}_{r'}(\vec{P}\,'))^*\nonumber\\
&&
\hspace{6.75cm}=\bar
u^{B'}_{r'}(\vec{P}\,')\gamma^0(\tilde\Gamma^{\lambda\,\mu}(P',P))^\dagger\gamma^0
{u}^{B}_{\lambda\,r}(\vec{P}),\nonumber\\
%
\tilde\Gamma^{\lambda\,\mu}(P',P)=&&
\left(-\frac{C_3^V(\omega)}{M'}(g^{\lambda\,\mu}q
\hspace{-.15cm}/\,
-q^\lambda\gamma^\mu)-\frac{C_4^V(\omega)}{M'^2}(g^{\lambda\,\mu}q\cdot P-q^\lambda
P^\mu)-\frac{C_5^V(\omega)}{M'^2}(g^{\lambda\,\mu}q\cdot P'-q^\lambda
P'^\mu)+C_6^V(\omega)g^{\lambda\,\mu}\right)\gamma_5\nonumber\\
&&+\left(-\frac{C_3^A(\omega)}{M'}(g^{\lambda\,\mu}q
\hspace{-.15cm}/\,
-q^\lambda\gamma^\mu)-\frac{C_4^A(\omega)}{M'^2}(g^{\lambda\,\mu}q\cdot P-q^\lambda
P^\mu)+{C_5^A(\omega)}g^{\lambda\,\mu}+\frac{C_6^A(\omega)}{M'^2}
q^\lambda q^\mu\right).
\end{eqnarray} 
Again, and within our model, we shall have that 
$C^V_{5}(\omega)=C^V_{6}(\omega)=C^A_{3}(\omega)=0$.

\item $\ 3/2\to3/2$ transitions.\\
A form factor decomposition for  $3/2\to3/2$ can be found in
Ref.~\cite{Faessler:2009xn} where a total of 7 vector plus 7 axial form factors
are needed. In this case we do not evaluate the form factors but work directly
with the vector and axial matrix elements.
\end{enumerate}
In appendix~\ref{app:ffwme} we give the expressions that relate the
form factors to weak current matrix elements and show how the latter
ones are evaluated in the model. Relations found between matrix
elements that simplify the calculation are also shown there.

\subsection{Results}
The results we obtain for the semileptonic decay widths of $cb$
baryons are presented in Tables~\ref{tab:resctos} ($c\to s$ decays)
and \ref{tab:resctod} ($c\to d$ decays).  We show between parentheses
the results obtained ignoring configuration mixing in the spin-1/2
$cb$ initial baryons. In this latter case, the
$\Xi^{(1)}_{cb},\ \Xi^{(2)}_{cb}$ baryons should be interpreted
respectively as the $\Xi'_{cb},\ \Xi_{cb}$ states of
Table~\ref{tab:baryons}.  We see small changes for transitions to
final states where the two light quarks couple to spin 0. On the other
hand, configuration mixing effects are very important for transitions
to final states where the two light quarks couple to spin 1, where we
find enhancements or reductions as large as a factor of 2.

\begin{table}[h!!!]
\begin{tabular}{llccc}
&\multicolumn{4}{c}{$\Gamma \ [10^{-14}\,{\rm GeV}]$}\\
&{This work}&\cite{sanchis95}&\cite{Faessler:2001mr}&\cite{Kiselev:2001fw}\\
\hline
$\Xi^{(1)\,+}_{cbu}\to\Xi^0_b\, e^+\nu_e$& 3.74 (3.45)&(3.4)\\
$\Xi^{(2)\,+}_{cbu}\to\Xi^0_b\, e^+\nu_e$& 2.65 (2.87)\\
$\Xi^{(1)\,+}_{cbu}\to\Xi'^0_b\, e^+\nu_e$& 3.88
(1.66)&&$2.44\div3.28^\dagger$\\
$\Xi^{(2)\,+}_{cbu}\to\Xi'^0_b\, e^+\nu_e$&1.95 (3.91)\\
$\Xi^{(1)\,+}_{cbu}\to\Xi^{*\,0}_b\, e^+\nu_e$& 1.52 (3.45)\\
$\Xi^{(2)\,+}_{cbu}\to\Xi^{*\,0}_b\, e^+\nu_e$& 2.67 (1.02)\\
$\Xi^{(2)\,+}_{cbu}\to\Xi^0_b\, e^+\nu_e+\Xi'^0_b\, e^+\nu_e+\Xi^{*\,0}_b\, e^+\nu_e$& 7.27 (7.80)
&&&$(9.7\pm1.3)^*$\\
$\Xi^{*\,+}_{cbu}\to\Xi^{0}_b\, e^+\nu_e$&  4.08\\
$\Xi^{*\,+}_{cbu}\to\Xi'^{0}_b\, e^+\nu_e$&0.747\\
$\Xi^{*\,+}_{cbu}\to\Xi^{*\,0}_b\, e^+\nu_e$& 5.03\\\hline
\end{tabular}\hspace{.5cm}
\begin{tabular}{ll}
&{\hspace*{.375cm}$\Gamma \ [10^{-14}\,{\rm GeV}]$}\\
\hline
$\Omega^{(1)\,0}_{cbs}\to\Omega^-_b\, e^+\nu_e$&\hspace*{.5cm} 7.21 (3.12)\\
$\Omega^{(2)\,0}_{cbs}\to\Omega^-_b\, e^+\nu_e$&\hspace*{.5cm} 3.49 (7.12)\\
$\Omega^{(1)\,0}_{cbs}\to\Omega^{*\,-}_b\, e^+\nu_e$&\hspace*{.5cm} 2.98 (6.90)\\
$\Omega^{(2)\,0}_{cbs}\to\Omega^{*\,-}_b\, e^+\nu_e$&\hspace*{.5cm} 5.50 (2.07)\\
$\Omega^{*\,0}_{cbs}\to\Omega^-_b\, e^+\nu_e$&\hspace*{.5cm} 1.35\\
$\Omega^{*\,0}_{cbs}\to\Omega^{*\,-}_b\, e^+\nu_e$&\hspace*{.5cm} 10.2\\\hline
\end{tabular}
\caption{$\Gamma$ decay widths for  $c\to s$ decays. Results where configuration
mixing is not considered are shown in between parentheses. The result with a
$\dagger$ corresponds to the decay of the $\widehat \Xi_{cb}$
state. The result 
with an ${\ast}$ is our
estimate from the total decay width and the branching ratio
given in~\cite{Kiselev:2001fw}. Similar results are obtained for decays into
$\mu^+\nu_\mu$.} 
\label{tab:resctos}
\end{table}
\begin{table}[h!!!]
\begin{tabular}{ll}
&{\hspace*{.5cm}$\Gamma \ [10^{-14}\,{\rm GeV}]$}\\
\hline
$\Xi^{(1)\,+}_{cbu}\to\Lambda^0_b\, e^+\nu_e$&\hspace*{.5cm} 0.219 (0.196)\\
$\Xi^{(2)\,+}_{cbu}\to\Lambda^0_b\, e^+\nu_e$&\hspace*{.5cm}  0.136 (0.154)\\
$\Xi^{(1)\,+}_{cbu}\to\Sigma^0_b\, e^+\nu_e$&\hspace*{.5cm}  0.198 (0.0814)\\
$\Xi^{(2)\,+}_{cbu}\to\Sigma^0_b\, e^+\nu_e$ &\hspace*{.5cm} 0.110 (0.217)\\
$\Xi^{(1)\,+}_{cbu}\to\Sigma^{*\,0}_b\, e^+\nu_e$&\hspace*{.5cm}  0.0807 (0.184)\\
$\Xi^{(2)\,+}_{cbu}\to\Sigma^{*\,0}_b\, e^+\nu_e$&\hspace*{.5cm}  0.147 (0.0556)\\
$\Xi^{*\,+}_{cbu}\to\Lambda^{0}_b\, e^+\nu_e$&\hspace*{.5cm}   0.235\\
$\Xi^{*\,+}_{cbu}\to\Sigma^{0}_b\, e^+\nu_e$&\hspace*{.5cm}  0.0399\\
$\Xi^{*\,+}_{cbu}\to\Sigma^{*\,0}_b\, e^+\nu_e$&\hspace*{.5cm}  0.246\\\hline
\end{tabular}\hspace{2cm}
\begin{tabular}{ll}
&{\hspace*{.5cm}$\Gamma \ [10^{-14}\,{\rm GeV}]$}\\
\hline
$\Omega^{(1)\,0}_{cbs}\to\Xi^-_b\, e^+\nu_e$&\hspace*{.5cm}  0.179 (0.164)\\
$\Omega^{(2)\,0}_{cbs}\to\Xi^-_b\, e^+\nu_e$&\hspace*{.5cm}  0.120 (0.133)\\
$\Omega^{(1)\,0}_{cbs}\to\Xi'^-_b\, e^+\nu_e$&\hspace*{.5cm}  0.169 (0.0702)\\
$\Omega^{(2)\,0}_{cbs}\to\Xi'^-_b\, e^+\nu_e$&\hspace*{.5cm}  0.0908 (0.182)\\
$\Omega^{(1)\,0}_{cbs}\to\Xi^{*\,-}_b\, e^+\nu_e$&\hspace*{.5cm}  0.0690 (0.160)\\
$\Omega^{(2)\,0}_{cbs}\to\Xi^{*\,-}_b\, e^+\nu_e$&\hspace*{.5cm}  0.130 (0.0487)\\
$\Omega^{*\,0}_{cbs}\to\Xi^-_b\, e^+\nu_e$&\hspace*{.5cm}  0.196\\
$\Omega^{*\,0}_{cbs}\to\Xi'^-_b\, e^+\nu_e$&\hspace*{.5cm}  0.0336\\
$\Omega^{*\,0}_{cbs}\to\Xi^{*\,-}_b\, e^+\nu_e$&\hspace*{.5cm}  0.223\\\hline
\end{tabular}\caption{$\Gamma$ decay widths for  
$c\to d$ decays. In between parentheses
we show the results without configuration mixing. Similar results are obtained
for decays into
$\mu^+\nu_\mu$.}
\label{tab:resctod}
\end{table}
Note also that, even though 
$|V_{cs}|^2,\,|V_{cd}|^2\gg|V_{cb}|^2$, the values we get for the  decay 
widths are of the same order of magnitude
 to what we obtained for $b\to c$ transitions in 
Ref.~\cite{Albertus:2009ww}. In the present
case, the greater
value of the CKM matrix elements are compensated by a smaller phase space.

In the left panel of Table~\ref{tab:resctos} we compare our results to
the few available results obtained by other groups (we have not found
in the literature any previous result for $c\to d$ decays to compare
with our predictions in Table~\ref{tab:resctod}).  Our estimate,
without configuration mixing, for the $\Xi^{(1)}_{cb}\to\Xi_b$
transition agrees very well with the one obtained in
Ref.~\cite{sanchis95}.  For the $\Xi^{(1)\,+}_{cbu}\to\Xi'^0_b$
transition we are also in agreement with the calculation in
Ref.~\cite{Faessler:2001mr}. There, the authors use the
$\widehat\Xi_{cb}$ baryon which is almost equal to our physical state
$\Xi^{(1)}_{cb}$.  We also see that our result for the combined decay
$(\Xi^{(1)\,+}_{cbu}\to\Xi^0_b)+(\Xi^{(1)\,+}_{cbu}\to\Xi^{\prime\,0}_b)+(
\Xi^{(1)\,+}_{cbu}\to\Xi^{*\,0}_b)$ is in
reasonable agreement with the one predicted in
Ref.~\cite{Kiselev:2001fw}. This combined decay width is not very
sensitive to configuration mixing effects.

Besides the results shown in Tables~\ref{tab:resctos} and \ref{tab:resctod},
  we have from isospin symmetry that
\bea
\Gamma(\Xi_{cbd}^{(1)\,0}\to\Sigma_b^-)\approx2\,\Gamma(\Xi_{cbu}^{(1)\,+}\to\Sigma_b^0)\ \ &,&\ \
\Gamma(\Xi_{cbd}^{(1)\,0}\to\Sigma_b^{*\,-})\approx2\,
\Gamma(\Xi_{cbu}^{(1)\,+}\to\Sigma_b^{*\,0}),\nonumber\\
\Gamma(\Xi_{cbd}^{(2)\,0}\to\Sigma_b^-)\approx
2\,\Gamma(\Xi_{cbu}^{(2)\,+}\to\Sigma_b^0)\ \ &,&\ \
\Gamma(\Xi_{cbd}^{(2)\,0}\to\Sigma_b^{*\,-})\approx2\,\Gamma(\Xi_{cbu}^
{(2)\,+}\to\Sigma_b^{*\,0}),\nonumber\\
\Gamma(\Xi_{cbd}^{*\,0}\to\Sigma_b^-)\approx
2\,\Gamma(\Xi_{cbu}^{*\,+}\to\Sigma_b^0)\ \ &,&\ \ 
\Gamma(\Xi_{cbd}^{*\,0}\to\Sigma_b^{*\,-})\approx2\,\Gamma(\Xi_{cbu}^
{*\,+}\to\Sigma_b^{*\,0}).\nonumber\\
\eea
\bea
\Gamma(\Xi_{cbd}^{(1)\,0}\to\Xi_b^-)\approx\Gamma(\Xi_{cbu}^{(1)\,+}\to\Xi_b^0)\ \ ,\ \
&\Gamma(\Xi_{cbd}^{(1)\,0}\to\Xi_b^{\prime\,-})\approx
\Gamma(\Xi_{cbu}^{(1)\,+}\to\Xi_b^{\prime\,0})&\ \ ,\ \ 
\Gamma(\Xi_{cbd}^{(1)\,0}\to\Xi_b^{*\,-})\approx
\Gamma(\Xi_{cbu}^{(1)\,+}\to\Xi_b^{*\,0}),\nonumber\\
\Gamma(\Xi_{cbd}^{(2)\,0}\to\Xi_b^-)\approx\Gamma(\Xi_{cbu}^{(2)\,+}\to\Xi_b^0)\ \ ,\ \
&\Gamma(\Xi_{cbd}^{(2)\,0}\to\Xi_b^{\prime\,-})\approx
\Gamma(\Xi_{cbu}^{(2)\,+}\to\Xi_b^{\prime\,0})&\ \ ,\ \ 
\Gamma(\Xi_{cbd}^{(2)\,0}\to\Xi_b^{*\,-})\approx
\Gamma(\Xi_{cbu}^{(2)\,+}\to\Xi_b^{*\,0}),\nonumber\\
\Gamma(\Xi_{cbd}^{*\,0}\to\Xi_b^-)\approx\Gamma(\Xi_{cbu}^{*\,+}\to\Xi_b^0)\ \ ,\ \
&\Gamma(\Xi_{cbd}^{*\,0}\to\Xi_b^{\prime\,-})\approx
\Gamma(\Xi_{cbu}^{*\,+}\to\Xi_b^{\prime\,0})&\ \ ,\ \ 
\Gamma(\Xi_{cbd}^{*\,0}\to\Xi_b^{*\,-})\approx
\Gamma(\Xi_{cbu}^{*\,+}\to\Xi_b^{*\,0}).\nonumber\\
\eea
The sources of uncertainties in the present calculation are the same
as the ones we discussed for the $c\to s,d$ decays of $cc$ baryons in
Ref.~\cite{Albertus:2011xz}.  First, the use of different interquark
potentials, like the AP1~\cite{semay94,silvestre96} and
Bhaduri~\cite{BD81} potentials, to evaluate the wave functions could
change  the decay widths at the level of $10\,$\%. This can be
considered as part of the uncertainties inherent to our model.
Another important source of
uncertainties is our lack of knowledge of the actual masses of the
$cb$ baryons.  For instance,  a reduction of $70\,$MeV
in the $\Omega^*_{cb}$ mass (a mere 1\% reduction) makes de
$\Omega^*_{cb}\to\Omega'_b$ decay width smaller by some 25\%.
Precise decay widths predictions should await for a precise mass
knowledge of $cb$ baryons. Moreover, one has the possible contribution of 
intermediate $D^*$ and
$D^*_s$ vector meson exchanges~\cite{Isgur:1989qw,Albertus:2005ud}. 
This mechanisms
are not considered in our calculation neither have they been taken
into account in the previous ones of Refs.~\cite{sanchis95,Faessler:2001mr,
  Kiselev:2001fw}. We expect such exchanges to produce small effects
as the $D^*$ and $D^*_s$ poles are located far from $\sqrt{q^2_{\rm
    max}}$. In any case, with the intermediate vector mesons being far
off-shell, the computation of their effects will be complicated due to
the unknown strength of their couplings with the singly and doubly
heavy baryons, and the lack of a reasonable scheme to model how the
latter interactions are suppressed when $q^2$ approaches the endpoint
of the available phase-space ($q^2=0$). 
From our experience in the
previous work of Ref.~\cite{Albertus:2005ud}, in particular from 
 the $D\to K$ semileptonic decay  where similar $q^2$
exchanges were involved, we would expect  vector  meson exchange  effects
in the decay widths  to be  below  the 25\% already mentioned above. 
\section{Heavy quark spin symmetry}
\label{sect:hqss}
 In this section we use HQSS to derive model independent, though approximate,
relations between different form factors and  decay widths. 
This is similar to what we did for $b\to c$ decays of $cb$ baryons in Ref.~\cite{Hernandez:2007qv}
or more recently for  $b\to c$ transitions of triply heavy 
baryons in Ref.~\cite{Flynn:2011gf}.

 The
consequences of spin symmetry for weak matrix elements can be derived
using the ``trace formalism''~\cite{Falk:1990yz,MWbook}. To represent
the lowest-lying $S$-wave $cb$ baryons we will use wave-functions
made of tensor products of Dirac matrices and spinors,
namely~\cite{Flynn:2007qt}:
\bea
\widehat B_{cb}&\to&{\left[\frac{1+/\hspace{-.175cm}
v}{2}\gamma_\lambda\right]_{\alpha\beta}}
{\left[\frac1{\sqrt3}(v^\lambda+\gamma^\lambda)\gamma_5u(v,r)\right]_\gamma},
\nonumber\\
\widehat B'_{cb}&\to&{\left[-\frac{1+/\hspace{-.175cm} v}{2}\gamma_5
\right]_{\alpha\beta}}
{u_\gamma(v,r)},\nonumber\\
\widehat B^*_{cb}&\to&{\left[\frac{1+/\hspace{-.175cm} v}{2}\gamma_\lambda\right]
_{\alpha\beta}}
{u^\lambda_\gamma(v,r)},
\eea
where we have indicated Dirac indices $\alpha$, $\beta$ and $\gamma$
explicitly on the right-hand side and $r$ is a helicity label for the
baryon. These wave functions describe states\footnote{States are
  normalized to $-2 M = -\bar u u=\bar u^\lambda u_\lambda$.} where
the $c$ quark and the light quark couple to definite spin 0 ($\widehat
B'_{cb}$) or 1 ($\widehat B_{cb},\,\widehat B^*_{cb}$). The $b$ quark
couples with that subsystem to total spin 1/2 ($\widehat
B_{cb},\,\widehat B'_{cb}$) or 3/2 ($\widehat B^*_{cb}$). Note that
$\widehat B^*_{cb}=B^*_{cb}$. Under a Lorentz transformation,
$\Lambda$, and quark spin rotations $S_c$ and $S_b$ for $c$ and $b$
quarks a wave-function of the form $\Gamma_{\alpha\beta}\, {\cal
  U}_\gamma$ transforms as:
\begin{equation}
\label{eq:spintransfs}
\Gamma_{\alpha\beta}\,{\cal U}_\gamma \to \left[S(\Lambda) \Gamma 
S^{-1}(\Lambda)\right]_{\alpha\beta}\; \left[S(\Lambda){\cal U}\right]_\gamma,
\quad
\Gamma_{\alpha\beta}\,{\cal U}_\gamma \to \left[S_c \Gamma\right]_{\alpha\beta}
\, \left[S_b\, {\cal U}\right]_\gamma.
\end{equation}
with ${\cal U} = u, \frac1{\sqrt3}(v^\lambda+\gamma^\lambda)\gamma_5
u, u^\lambda$. On the other hand, the final $b$ baryons 
 are represented by the following spinor wave
functions~\cite{MWbook}
\begin{align}
\label{eq:Lambda_c}
\Lambda_{b},\Xi_b &\to u'_\gamma (v',r'), \\
\Sigma_{b},\,\Xi'_b,\,\Omega_b &\to \left[\frac{1}{\sqrt3} (v^{\prime \lambda} + \gamma^\lambda)
  \g5  u'(v',r')\right]_\gamma, \\
\Sigma^*_{b},\,\Xi^{*}_b,\,\Omega^*_b &\to u'^\lambda_\gamma (v',r'),
\end{align}
where here the states are normalized to $-2M'$.
In this case we have 
that
\begin{equation}
\label{eq:spintransfs2}
{\cal U}'_\gamma \to \left[ S(\Lambda){\cal U}'\right]_\gamma,
\quad
{\cal U}'_\gamma \to \left[  S_b\, {\cal U}\right ]_\gamma.
\end{equation}

The semileptonic 
decays are driven by the current $J^\mu= \bar l \gamma^\mu
(1-\gamma_5)c $, with $l=d,s$.  Under a $c$ quark spin rotation, it
transforms as $J^\mu \to J^\mu S_c^\dagger$. Thus, the only possible
amplitude that it is invariant under separate bottom and charm quark
spin rotations is of  the form 
\bea \bar {\cal U}'\, {\cal U}\ {\rm Tr}
\left[\gamma^\mu(1-\gamma_5)\Gamma\Omega\right], 
\eea 
where $\Omega$ is one of the two following functions, depending on whether the spin of
the light degrees of freedom in the final baryon ($S'_{\rm light}$) is
$0$ or $1$ 
\bea
\Omega &=&\eta_1+\eta_2/\hspace{-.15cm}v', \qquad {\rm for}~~ S'_{\rm
  light} = 0\nonumber\\ 
\Omega_\lambda &=&\beta_1\gamma_\lambda+\beta_2/\hspace{-.15cm}v'\gamma_\lambda+
\beta_3 v_\lambda+\beta_4/\hspace{-.15cm}v'v_\lambda, \qquad {\rm for}~~ S'_{\rm
  light} = 1
\eea 
Terms in $/\hspace{-.15cm}v$ are not included since
$\frac{1+/\hspace{-.15cm}v}2/\hspace{-.15cm}v=\frac{1+/\hspace{-.15cm}v}2$.
We are interested in the transition matrix elements close to zero recoil 
where we have that $v'^\mu\approx v^\mu$, $\bar u'\gamma_5 u\approx0$,
$v^\mu\bar u'\bar u\approx
v'^\mu\bar u'\bar u\approx\bar u'\gamma^\mu\bar u$. Besides we have the exact
relations
\bea
v_\lambda u^\lambda=v_\lambda(v^\lambda+\gamma^\lambda)\gamma_5u=0,\nonumber\\
\bar u^{\prime\,\lambda} v'_\lambda=\bar
u^{\prime\,\lambda}\gamma_5(v^{\prime\lambda}+\gamma^\lambda)v'_\lambda=0,\nonumber\\
/\hspace{-.175cm} vu=u, \bar u'/\hspace{-.175cm} v'=\bar u',\nonumber\\
\gamma_\lambda u^\lambda=
\bar u^{\prime\,\lambda}\gamma_\lambda=0.
\eea
Taking into account all this, we can obtain approximate expressions
for the hadronic matrix elements that are valid near the zero recoil
point. Apart from 
global phases we get the following results:
\begin{itemize}
\item $\widehat B_{cb}\to\Lambda_b,\Xi_b$

\bea
&&\bar u'\frac1{\sqrt3}\big(v^\sigma+\gamma^\sigma\big)\gamma_5u
\ {\rm Tr}\big[\gamma^\mu(1-\gamma_5)\frac{1+/\hspace{-.175cm} v}{2}\gamma_\sigma
\Omega\big]
\approx\frac2{\sqrt3}\big(\eta_1-\eta_2\big)
\bar u'\gamma^\mu\gamma_5u
=\frac1{\sqrt3}\eta\,
\bar u'\big(-\gamma^\mu\gamma_5\big)u,
\label{eq:hqss1}
\eea 
where we have introduced $\eta=-2(\eta_1-\eta_2)$. This is a
function that depends only on $\omega$, and it is the analog of the
Isgur-Wise function firstly introduced in the context of $b \to c$
semileptonic meson
decays~\cite{MWbook}. 

  We see that near the zero recoil point, HQSS
considerably reduces the number of independent form factors. In fact
we find  that for $\omega=1$,
\bea
F_1+F_2+F_3=0\ ,\ \ G_1=\frac1{\sqrt3}\eta. 
\eea
\item $\widehat B'_{cb}\to \Lambda_b,\Xi_b$
\bea
\bar u'u
\ {\rm Tr}\big[\gamma^\mu(1-\gamma_5)(-1)\frac{1+/\hspace{-.175cm} v}{2}\gamma_5
\Omega\big]\approx-2\big(\eta_1-\eta_2\big)
\bar u'\gamma^\mu u=\eta\,\bar u'\gamma^\mu u,
\label{eq:hqss2}
\eea
from where one can conclude that at $\omega=1$
\bea
F_1+F_2+F_3=\eta\ ,\ \ G_1=0.
\eea

\item  $\widehat B^*_{cb}\to \Lambda_b, \Xi_b$ 
\bea
\bar u'u^\sigma
\ {\rm Tr}\big[\gamma^\mu(1-\gamma_5)\frac{1+/\hspace{-.175cm}
v}{2}\gamma_\sigma
\Omega\big]\approx2\big(\eta_1-\eta_2\big)
\bar u^{\prime\,} u^\mu=-\eta\,\bar u^{\prime\,} u^\mu,
\label{eq:hqss3}
\eea
which in this case implies that at $\omega=1$
\bea
-C_3^A\frac{M-M'}{M'}-C_4^A\frac{M(M-M')}{M'^2}+C_5^A=-\eta.
\eea  
\end{itemize}
The $\eta$ Isgur-Wise function is different for different light
quark configurations in the final state and depends also on whether
the initial light quark is an $n=u,d$ quark or a $s$ quark. However,
$SU(3)$ flavor symmetry could be used to establish relations between
all of them.  Besides $\eta$ would be normalized to 1 at zero recoil
($\eta(1)=1$) in the equal mass case. In the actual calculation
deviations from this limiting value are expected due to the mismatch
of the initial and final baryons wave functions.

\begin{itemize}
\item $\ \widehat B_{cb}\to \Sigma_b, \Xi'_b, \Omega_b$ 

\bea
-\bar u'\frac1{\sqrt3}\gamma_5\big(v^{\prime\,\lambda}+\gamma^\lambda\big)\frac1{\sqrt3}\big(v^\sigma+\gamma^\sigma\big)\gamma_5u
\ {\rm Tr}\big[\gamma^\mu(1-\gamma_5)\frac{1+/\hspace{-.175cm} v}{2}\gamma_\sigma
\Omega_\lambda\big]&\approx&-2\big(\beta_1-\beta_2\big)
\bar u'\big(\gamma^\mu-\frac23\gamma^\mu\gamma_5\big)u\nonumber\\
&=&
\beta\bar u'\big(\gamma^\mu-\frac23\gamma^\mu\gamma_5\big)u,
\label{eq:hqss4}
\eea
where we have defined $\beta=-2(\beta_1-\beta_2)$, which is the Isgur-Wise
function in this case. For $\omega=1$ one would then obtain that
\bea
F_1+F_2+F_3=\beta ,\ \ G_1=\frac23\beta.
\eea
\item $\ \widehat B'_{cb}\to \Sigma_b, \Xi'_b, \Omega_b$
\bea
-\frac1{\sqrt3}\bar u'\gamma_5\big(v^{\prime\,\lambda}+\gamma^\lambda\big)
u
\ {\rm Tr}\big[\gamma^\mu(1-\gamma_5)(-1)\frac{1+/\hspace{-.175cm} v}{2}
\gamma_5
\Omega_\lambda\big]&\approx&\frac2{\sqrt3}\big(\beta_1-\beta_2\big)
\bar u'\gamma^\mu\gamma_5u
=\frac1{\sqrt3}\beta\,
\bar u'\big(-\gamma^\mu\gamma_5\big)u,
\label{eq:hqss5}
\eea
that for $\omega=1$ implies that
\bea
F_1+F_2+F_3=0,\ \ G_1=\frac1{\sqrt3}\beta.
\eea
\item $\ \widehat B^*_{cb}\to \Sigma_b, \Xi'_b, \Omega_b$
\bea
-\bar u'\frac1{\sqrt3}\gamma_5\big(v^{\prime\,\lambda}+\gamma^\lambda\big)
u^\sigma
\ {\rm Tr}\big[\gamma^\mu(1-\gamma_5)\frac{1+/\hspace{-.175cm} v}{2}\gamma_\sigma
\Omega_\lambda\big]\approx-\frac{2}{\sqrt3}(\beta_1-\beta_2)\bar u'u^\mu
=\frac{1}{\sqrt3}\beta\,\bar u'u^\mu,
\label{eq:hqss6}
\eea
from where at $\omega=1$
\bea
-C_3^A\frac{M-M'}{M'}-C_4^A\frac{M(M-M')}{M'^2}+C_5^A=\frac1{\sqrt3}\beta.
\eea
\item $\ \widehat B_{cb}\to \Sigma^*_b, \Xi^*_b, \Omega^*_b$
\bea
\bar u^{\prime\lambda}\frac1{\sqrt3}\big(v^\sigma+\gamma^\sigma\big)\gamma_5u
\ {\rm Tr}\big[\gamma^\mu(1-\gamma_5)\frac{1+/\hspace{-.175cm} v}{2}\gamma_\sigma
\Omega_\lambda\big]\approx-\frac{2}{\sqrt3}\big(\beta_1-\beta_2\big)
\bar u^{\prime\mu}u=\frac{1}{\sqrt3}\beta\,
\bar u^{\prime\mu}u,
\label{eq:hqss7}
\eea
and thus at $\omega=1$ we have
\bea
C_3^A\frac{M-M'}{M}+C_4^A\frac{M'(M-M')}{M^2}+C_5^A=\frac1{\sqrt3}\beta.
\eea
\item $\ \widehat B'_{cb}\to \Sigma^*_b, \Xi^*_b, \Omega^*_b$
\bea
&&\bar u^{\prime\,\lambda}u
\ {\rm Tr}\big[\gamma^\mu(1-\gamma_5)(-1)\frac{1+/\hspace{-.175cm} v}{2}\gamma_5
\Omega_\lambda\big]
\approx2\big(\beta_1-\beta_2\big)\bar u^{\prime\,\mu}u=-\beta\bar u^{\prime\,\mu}u.
\label{eq:hqss8}
\eea
One obtains in this case that at $\omega=1$ 
\bea
C_3^A\frac{M-M'}{M}+C_4^A\frac{M'(M-M')}{M^2}+C_5^A=-\beta.
\eea
\item $\ \widehat B^*_{cb}\to \Sigma^*_b, \Xi^*_b, \Omega^*_b$
\bea
\bar u^{\prime\,\lambda}u^\sigma
\ {\rm Tr}\big[\gamma^\mu(1-\gamma_5)\frac{1+/\hspace{-.175cm} v}{2}\gamma_\sigma
\Omega_\lambda\big]\approx2\big(\beta_1 -\beta_2\big)
\bar u^{\prime\,\lambda}\gamma^\mu(1-\gamma_5)u_\lambda=
-\beta\,\bar u^{\prime\,\lambda}\gamma^\mu(1-\gamma_5)u_\lambda,
\label{eq:hqss9}
\eea
which implies for instance that the $V^0$ vector matrix element should be equal
to $-\beta$ at $\omega=1$ when evaluated in between states with the same spin
projection.
\end{itemize}
As for the $\eta$ function above, the $\beta$ Isgur-Wise function is different
 for different light quark 
configurations in the final state and depends also 
on whether the initial light quark is an $n=u,d$ quark or a $s$ quark. 
Besides, if the quarks involved in the weak decay had equal mass  one would
 have that
$\beta(1)=1$ when the two light quarks in the final baryon are different 
($\Sigma^0_b,\, \Sigma^{*\,0}_b,\,
\Xi'^0_b,\,\Xi^{*\,0}_b ,\,\Xi'^-_b,\,\Xi^{*\,-}_b$) and  $\beta(1)=\sqrt2$ 
when they are identical ($\Sigma^-_b,\, \Sigma^{*\,-}_b,\,
\Omega_b^-,\, \Omega_b^{*\,-}$). 
Again, in the actual calculation deviations from these 
limiting values are
expected due to the mismatch of the initial and final baryon wave functions.

\begin{figure}
\rotatebox{270}{\resizebox{!}{12cm}{\includegraphics{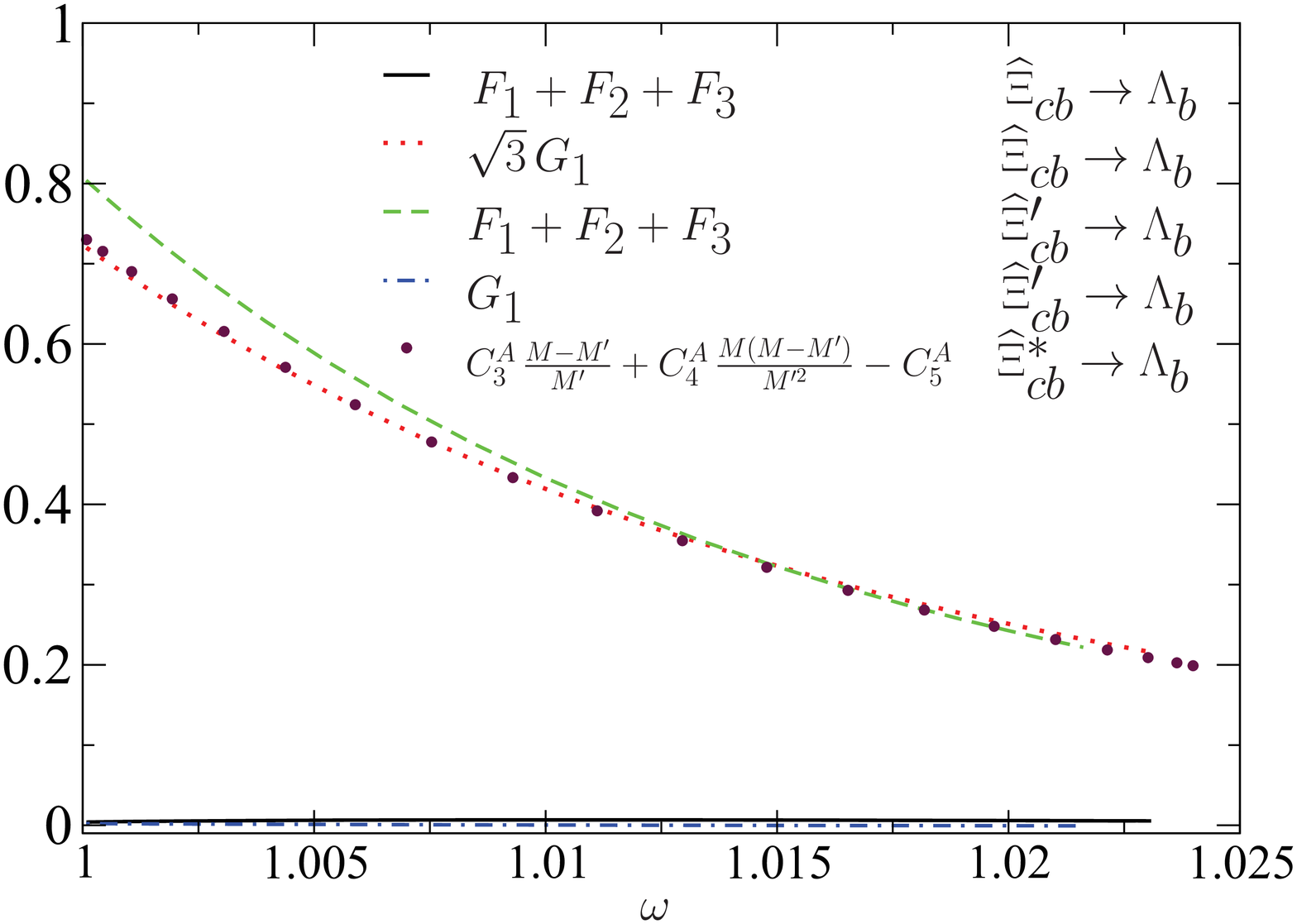}}}
\caption{Test of HQSS constraints: Different combinations of form
  factors obtained in this work for several transitions with a
  $\Lambda_b$ in the final state ($S'_{\rm light}=0$).  For the
  calculation we have taken the masses of the
  $\widehat\Xi_{cb},\widehat\Xi'_{cb}$ to be the masses of the
  physical states $\Xi^{(1)}_{cb},\Xi^{(2)}_{cb}$.  Similar results
  are obtained for the $\widehat\Omega_{cb},\widehat\Omega'_{cb},
  \widehat\Omega^*_{cb}\to\Xi_b$ and the
  $\widehat\Xi_{cb},\widehat\Xi'_{cb},\widehat\Xi^*_{cb}\to\Xi_b$
  transitions.\vspace*{.5cm}}
\label{fig:xitola}
\end{figure}

\begin{figure}
\rotatebox{270}{\resizebox{!}{12cm}{\includegraphics{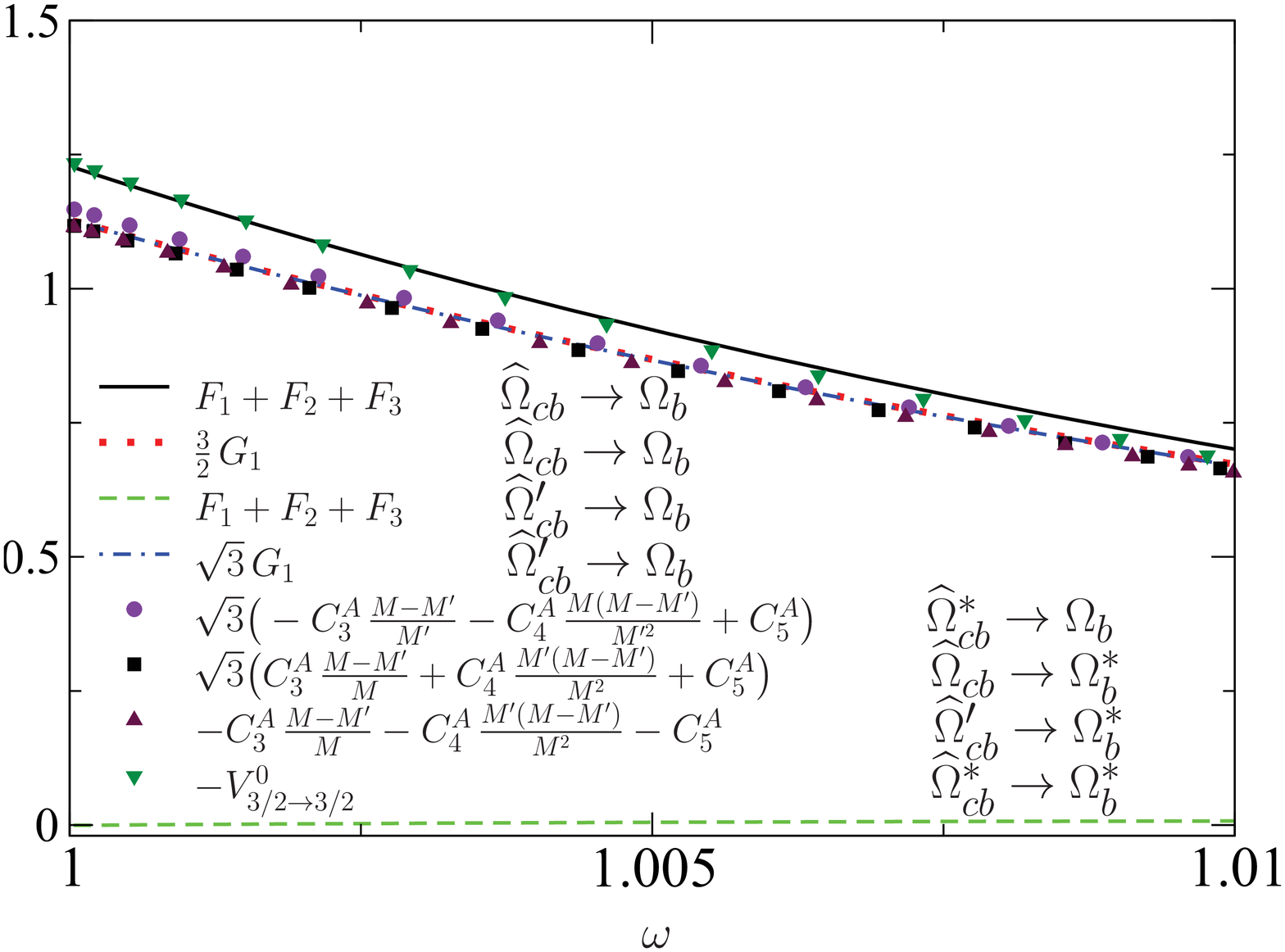}}}
\caption{Test of HQSS constraints: Different combinations of form
  factors obtained in this work for transitions with a
  $\Omega_b,\,\Omega^*_b$ in the final state ($S'_{\rm
    light}=1$). $V^0_{3/2\to3/2}$ stands for the matrix element of the
  zero component of the vector current for spin projections 3/2 both
  in the initial and final baryon. For the calculation we have taken
  the masses of the $\widehat\Omega_{cb},\widehat\Omega'_{cb}$ to be
  the masses of the physical states
  $\Omega^{(1)}_{cb},\Omega^{(2)}_{cb}$.  Similar results are obtained
  for the
  $\widehat\Xi_{cb},\widehat\Xi'_{cb},\widehat\Xi^*_{cb}\to\Sigma_b,\Sigma^*_b$,
  the $\widehat\Xi_{cb},\widehat\Xi'_{cb},\widehat\Xi^*_{cb}\to\Xi'_b,\Xi^*_b$,
  and the $\widehat\Omega_{cb},\widehat\Omega'_{cb},
  \widehat\Omega^*_{cb}\to\Xi'_b,\Xi^*_b$ transitions.}
\label{fig:omtoom}
\end{figure}

In Figs.~\ref{fig:xitola} and \ref{fig:omtoom} we check that our
calculation respects the constraints on the form factors deduced from
HQSS. For that purpose we have assumed the $\widehat B_{cb},\widehat
B'_{cb}$ states have masses equal to that of the physical ones
$B^{(1)}_{cb},B^{(2)}_{cb}$ . One sees deviations, due to corrections
in the inverse of the heavy quark masses, at the 10\% level near zero
recoil. In fact the constraints are satisfied to that level of
accuracy over the whole $\omega$ range accessible in the decays. We
found similar deviations in our recent study of the $c\to s,d$ decays
of double charmed baryons in Ref.~\cite{Albertus:2011xz}, where we
explicitly showed these discrepancies tend to disappear when the mass
of the heavy quark is made arbitrarily large.  One also sees that at
our results for $\eta(1), \beta(1)$ are systematically smaller than
would be expected if the quarks participating in the transition had
equal masses.  This reduced value is due to the mismatch in the wave
functions due to the different masses of the initial ($c$) and final
($d$ or $s$) quarks involved in the transition.

The results of Figs.~\ref{fig:xitola} and \ref{fig:omtoom} show HQSS
is then a useful tool to understand the dynamics of the $c\to s,d$
decays of $cb$ baryons, as it was also the case for their CKM
suppressed $b\to c$
decays~\cite{Hernandez:2007qv,Albertus:2009ww}. We take
advantage of this fact and we  now use
the HQSS approximate hadronic amplitudes in Eqs.~(\ref{eq:hqss1}),
(\ref{eq:hqss2}), (\ref{eq:hqss3}), (\ref{eq:hqss4}),
(\ref{eq:hqss5}), (\ref{eq:hqss6}), (\ref{eq:hqss7}), (\ref{eq:hqss8})
and (\ref{eq:hqss9}) to obtain model independent, though approximate,
relations between different decay widths. With the use of those HQSS
amplitudes and the leptonic tensor in Eq.(\ref{eq:lt}) we obtain that
near zero recoil 
\bea \widehat B_{cb}\to \Lambda_b,\Xi_b\hspace{1cm}
{\cal L}^{\alpha\beta}H_{\alpha\beta}&\approx&\frac{2MM'}3\eta^2\bigg[-A(4+2\omega)+B\bigg(
2\frac{(v\cdot q)(v'\cdot q)}{q^2}-(\omega+1)\bigg)\bigg],\\
\widehat B'_{cb}\to \Lambda_b,\Xi_b\hspace{1cm}
{\cal L}^{\alpha\beta}H_{\alpha\beta}&\approx&{2MM'\eta^2}
\bigg[A(4-2\omega)+B\bigg(
2\frac{(v\cdot q)(v'\cdot q)}{q^2}-(\omega-1)\bigg)\bigg],\\
\widehat B^*_{cb}\to \Lambda_b,\Xi_b\hspace{1cm}
{\cal L}^{\alpha\beta}H_{\alpha\beta}&\approx&
\frac{2MM'}3\eta^2(\omega+1)\bigg[-3A+B
\bigg(\frac{(v'\cdot q)^2}{q^2}-1
\bigg)
\bigg],\\
\widehat B_{cb}\to \Sigma_b,\Xi'_b,\Omega_b\hspace{1cm}
{\cal L}^{\alpha\beta}H_{\alpha\beta}&\approx&2MM'\beta^2\bigg[A\frac19(20-26\omega)
+B\frac19\bigg(26\frac{(v\cdot q)(v'\cdot q)}{q^2}+(5-13\omega)\bigg)\bigg],\\
\widehat B'_{cb}\to \Sigma_b,\Xi'_b,\Omega_b\hspace{1cm}
{\cal L}^{\alpha\beta}H_{\alpha\beta}&\approx&\frac{2MM'}3\beta^2\bigg[-A(4+2\omega)+B\bigg(
2\frac{(v\cdot q)(v'\cdot q)}{q^2}-(\omega+1)\bigg)\bigg],\\
\widehat B^*_{cb}\to \Sigma_b,\Xi'_b,\Omega_b\hspace{1cm}
{\cal L}^{\alpha\beta}H_{\alpha\beta}&\approx&
\frac{2MM'}9\beta^2(\omega+1)\bigg[-3A+B
\bigg(\frac{(v\cdot q)^2}{q^2}-1
\bigg)
\bigg],\\
\widehat B_{cb}\to \Sigma^*_b,\Xi^*_b,\Omega^*_b\hspace{1cm}
{\cal L}^{\alpha\beta}H_{\alpha\beta}&\approx&
\frac{4MM'}9\beta^2(\omega+1)\bigg[-3A+B
\bigg(\frac{(v'\cdot q)^2}{q^2}-1
\bigg)
\bigg],\\
\widehat B'_{cb}\to \Sigma^*_b,\Xi^*_b,\Omega^*_b\hspace{1cm}
{\cal L}^{\alpha\beta}H_{\alpha\beta}&\approx&
\frac{4MM'}3\beta^2(\omega+1)\bigg[-3A+B
\bigg(\frac{(v'\cdot q)^2}{q^2}-1
\bigg)
\bigg],\\
\widehat B^*_{cb}\to \Sigma^*_b,\Xi^*_b,\Omega^*_b\hspace{1cm}
{\cal L}^{\alpha\beta}H_{\alpha\beta}&\approx&MM'\beta^2\bigg[-A\frac89 \omega\big(1+2\omega^2\big)
+B\frac29\bigg(\frac{(v\cdot q)(v'\cdot q)}{q^2}
(20+8\omega^2)-\omega\big(6+4\omega^2\big)\bigg)\bigg].\nonumber\\
\end{eqnarray}
We can now follow our work in Ref~\cite{Hernandez:2007qv} and, near zero recoil,
take $\omega\approx1$ and, because $v'\approx v$, 
also approximate
\begin{equation}
\frac{(v\cdot q)^2}{q^2}\approx\frac{(v'\cdot q)(v\cdot
  q)}{q^2}\approx\frac{(v'\cdot q)^2}{q^2}.
\label{eq:X}
\end{equation}
Besides, for a light lepton $e$ or $\mu$ we have that $B\approx -A$ near zero
recoil.

Using those approximations and denoting by $X$ the quantity in
Eq.(\ref{eq:X}) we arrive at the following approximate results valid near zero
recoil
\begin{eqnarray}
\widehat B_{cb}\to \Lambda_b,\Xi_b\hspace{1cm}
{\cal L}^{\alpha\beta}H_{\alpha\beta}&\approx&-\frac{4MM'}3\eta^2A\big(
2+X\big)
\label{eq:aprox1},\\
\widehat B'_{cb}\to \Lambda_b,\Xi_b\hspace{1cm}
{\cal L}^{\alpha\beta}H_{\alpha\beta}&\approx&{4MM'\eta^2}
A\big(1-X\big)\label{eq:aprox2},\\
\widehat B^*_{cb}\to \Lambda_b,\Xi_b\hspace{1cm}
{\cal L}^{\alpha\beta}H_{\alpha\beta}&\approx&
-\frac{4MM'}3\eta^2A\big(2+X
\big)\label{eq:aprox3},\\
\widehat B_{cb}\to \Sigma_b,\Xi'_b,\Omega_b\hspace{1cm}
{\cal L}^{\alpha\beta}H_{\alpha\beta}&\approx&\frac{4MM'}9\beta^2A
\big(1-13X\big)\label{eq:aprox4},\\
\widehat B'_{cb}\to \Sigma_b,\Xi'_b,\Omega_b\hspace{1cm}
{\cal
L}^{\alpha\beta}H_{\alpha\beta}&\approx&-\frac{4MM'}3\beta^2A\big(2+X\big)
\label{eq:aprox5},\\
\widehat B^*_{cb}\to \Sigma_b,\Xi'_b,\Omega_b\hspace{1cm}
{\cal L}^{\alpha\beta}H_{\alpha\beta}&\approx&
-\frac{4MM'}9\beta^2A\big(2+X\big)\label{eq:aprox6},\\
\widehat B_{cb}\to \Sigma^*_b,\Xi^*_b,\Omega^*_b\hspace{1cm}
{\cal L}^{\alpha\beta}H_{\alpha\beta}&\approx&
-\frac{8MM'}9\beta^2A\big(2+X\big)\label{eq:aprox7},\\
\widehat B'_{cb}\to \Sigma^*_b,\Xi^*_b,\Omega^*_b\hspace{1cm}
{\cal L}^{\alpha\beta}H_{\alpha\beta}&\approx&
-\frac{8MM'}3\beta^2A\big(2+X\big)\label{eq:aprox8},\\
\widehat B^*_{cb}\to \Sigma^*_b,\Xi^*_b,\Omega^*_b\hspace{1cm}
{\cal L}^{\alpha\beta}H_{\alpha\beta}&\approx&-\frac{4MM'}9\beta^2A
\big(1+14X\big).\label{eq:aprox9}
\end{eqnarray}

\begin{figure}
\begin{center}
\rotatebox{270}{\resizebox{!}{12cm}{\includegraphics{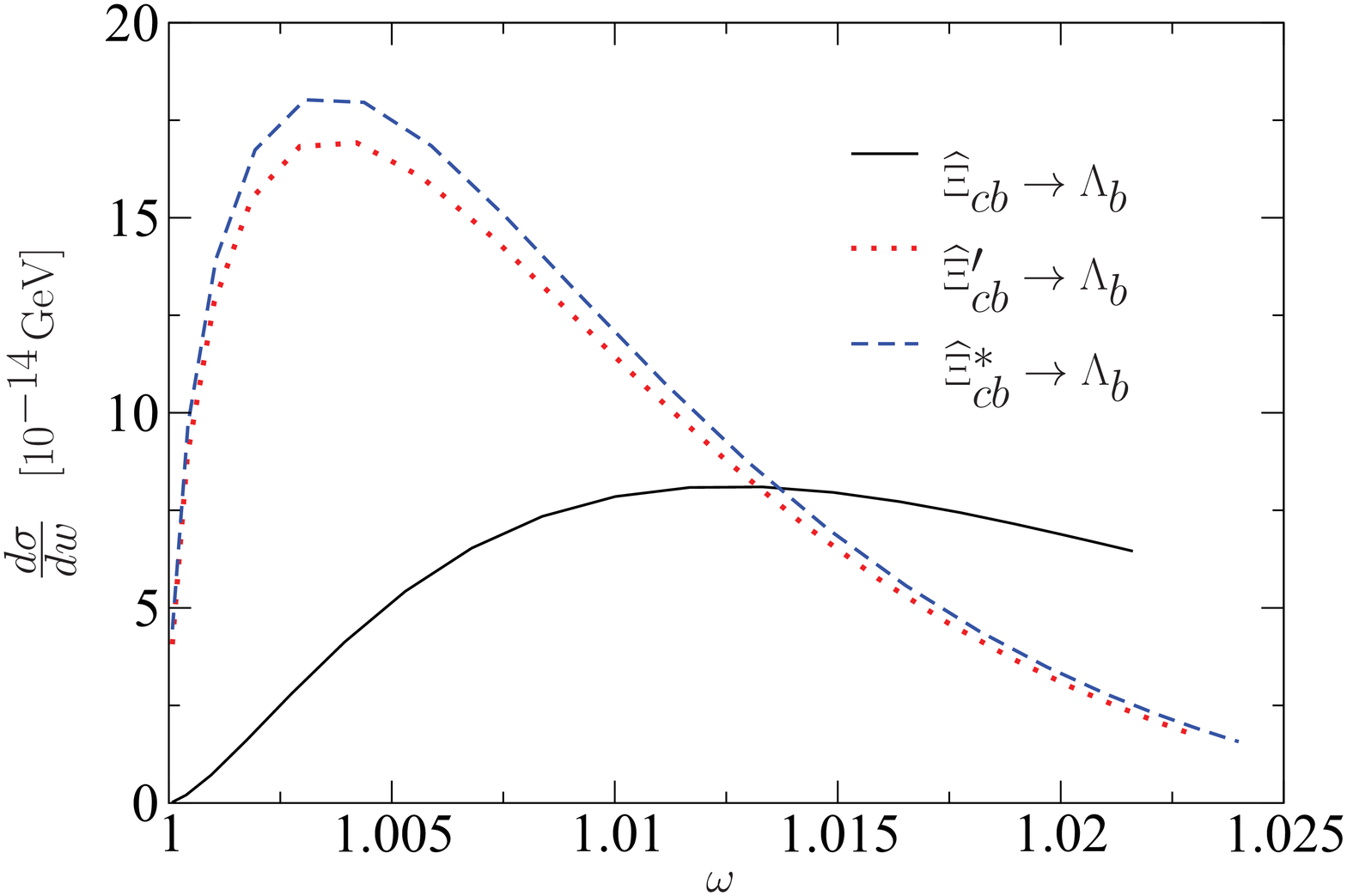}}}
\end{center}
\caption{Differential decay widths for the specified transitions.
\vspace*{.5cm}}
\label{fig:dsdw_lambda}
\end{figure}

\begin{figure}
\rotatebox{270}{\resizebox{!}{13cm}{\includegraphics{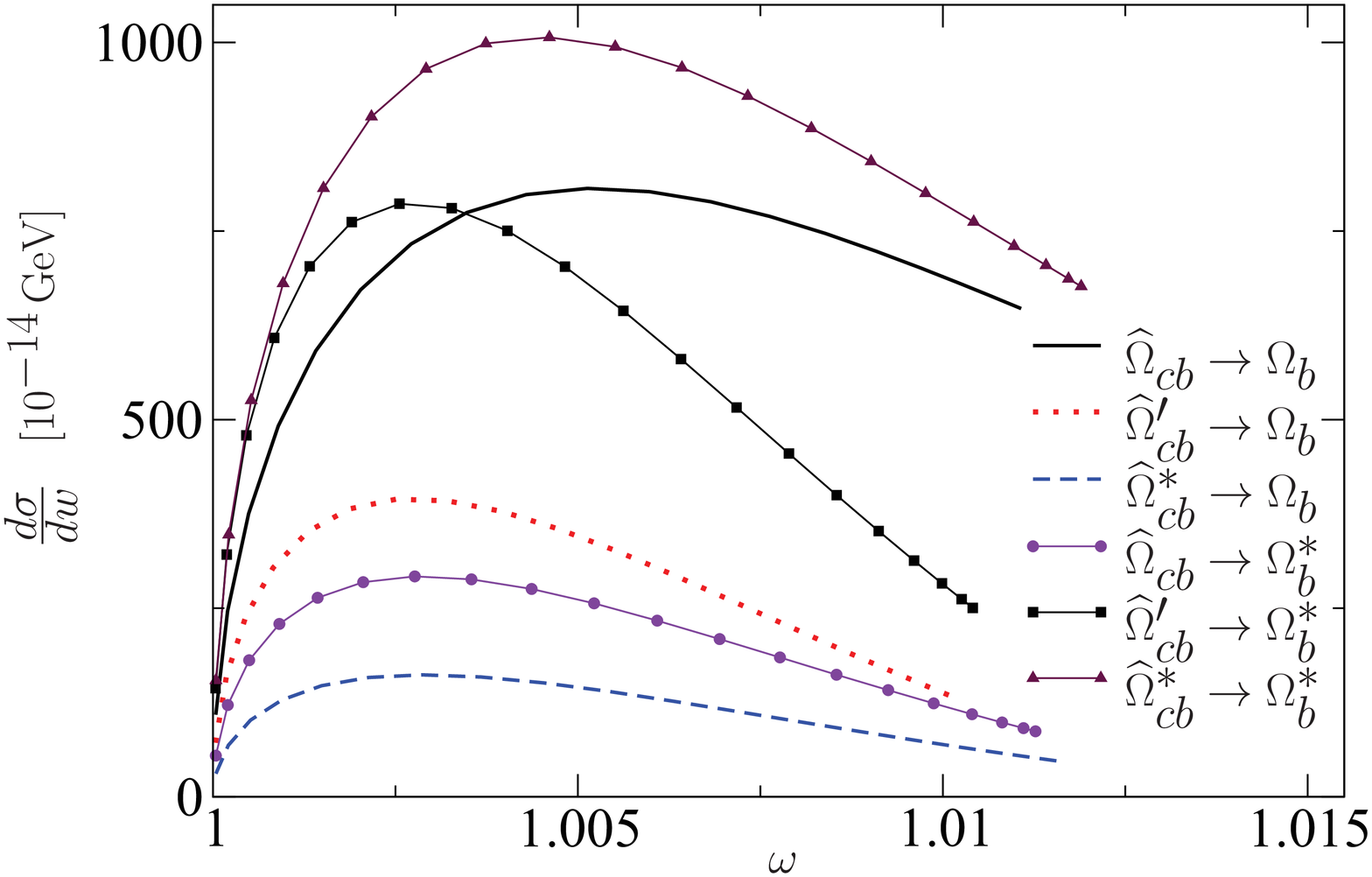}}}
\vspace*{.5cm}
\caption{Differential decay widths for the specified transitions.}
\label{fig:dsdw_omega}
\end{figure}

Can one extrapolate the above expressions over the whole $\omega$
range available in a given transition? In fact $B\approx -A$ to a very
high degree (better than one percent) practically in the whole
$\omega$ range accessible in these decays.  On the other hand
one has that $v\cdot q=M-M'\omega$, $v'\cdot q=M\omega-M'$ and one
expects larger deviations in approximate relation in Eq.~(\ref{eq:X})
for $\omega\approx \omega_{\rm max}$. For instance for the $\widehat
\Xi_{cb}\to\Lambda_b$ transition, one finds that $\frac{v'\cdot
  q}{v\cdot q}=1.20$ for $\omega=1+0.9(\omega_{\rm
  max}-1)$. Fortunately, the differential decay distributions peak at
much smaller $\omega$ values, so that errors related to the use of
Eq.~(\ref{eq:X}) in the whole $\omega$ range are less relevant.  We
show this in Figs.~\ref{fig:dsdw_lambda} and \ref{fig:dsdw_omega},
where we give differential decay widths for transitions with a
$\Lambda_b$ or an $\Omega^{(*)}_b$ in its final state. We have assumed
the masses of the $\widehat B_{cb},\widehat B'_{cb}$ to be the masses
of the physical states $B^{(1)}_{cb},B^{(2)}_{cb}$.


With this in mind and further assuming $M_{B_{cb}}= M_{B'_{cb}}= M_{B^*_{cb}}$ 
and 
$M_{B_{b}}= M_{B'_{b}}= M_{B^*_{b}}$ we can make  the following
approximate predictions based on HQSS

\bea
\Gamma(\widehat\Xi_{cb}\to\Lambda_b)&\approx&\Gamma(\widehat\Xi^*_{cb}\to\Lambda_b)
,\nonumber\\
\Gamma(\widehat B_{cb}\to\Xi_b)&\approx&\Gamma(\widehat B^*_{cb}\to\Xi_b),
\eea
\bea
\Gamma(\widehat\Xi'_{cb}\to\Sigma_b)&\approx&3\Gamma(\widehat\Xi^*_{cb}\to\Sigma_b)
\approx\frac32\Gamma(\widehat\Xi_{cb}\to\Sigma^*_b)\approx
\frac12\Gamma(\widehat\Xi'_{cb}\to\Sigma^*_b),\nonumber\\
\Gamma(\widehat B'_{cb}\to\Xi'_b)&\approx&3\Gamma(\widehat B^*_{cb}\to\Xi'_b)
\approx\frac32\Gamma(\widehat B_{cb}\to\Xi^*_b)\approx
\frac12\Gamma(\widehat B'_{cb}\to\Xi^*_b),\nonumber\\
\Gamma(\widehat\Omega'_{cb}\to\Omega_b)&\approx&3\Gamma(\widehat\Omega^*_{cb}\to\Omega_b)
\approx\frac32\Gamma(\widehat\Omega_{cb}\to\widehat\Omega^*_b)\approx
\frac12\Gamma(\widehat\Omega'_{cb}\to\Omega^*_b),
\eea
\bea
\Gamma(\widehat\Xi^*_{cb}\to\Sigma^*_b)&\approx&
\Gamma(\widehat\Xi^*_{cb}\to\Sigma_b)+\Gamma(\Xi_{cb}\to\Sigma_b),\nonumber\\
\Gamma(\widehat B^*_{cb}\to\Xi^*_b)&\approx&
\Gamma(\widehat B^*_{cb}\to\Xi'_b)+\Gamma(B_{cb}\to\Xi'_b),\nonumber\\
\Gamma(\widehat\Omega^*_{cb}\to\Omega^*_b)&\approx&
\Gamma(\widehat\Omega^*_{cb}\to\Omega_b)+
\Gamma(\widehat\Omega_{cb}\to\Omega_b).
\eea

Assuming that the states $\widehat B_{cb},\widehat B'_{cb}$ have the same masses as
the physical states $B^{(1)}_{cb},B^{(2)}_{cb}$ we get  the following 
numerical results
 (we give $\widetilde\Gamma=\frac{\Gamma}{10^{-14}\,{\rm GeV}}$)
\bea
\widetilde\Gamma(\widehat \Xi^{+}_{cb}\to\Lambda_b^0)&\approx&\widetilde\Gamma(\widehat \Xi^{*\,+}_{cb}\to\Lambda_b^0)
\nonumber\\0.219&\approx&0.235,
\label{eq:rel1}\\\nonumber\\
\widetilde\Gamma(\widehat \Omega^{0}_{cb}\to\Xi^-_b)&\approx&\widetilde\Gamma(\widehat \Omega^{*\,0}_{cb}\to\Xi^-_b)
\nonumber\\  0.179&\approx& 0.196,
\label{eq:rel2}\\\nonumber\\
\widetilde\Gamma(\widehat \Xi^{\,+}_{cb}\to\Xi^0_b)&\approx&\widetilde\Gamma(\widehat \Xi^{*\,+}_{cb}\to\Xi^0_b)
\nonumber\\   3.74&\approx& 4.08,  
\label{eq:rel3}\eea
\bea
\widetilde\Gamma(\widehat \Xi^{\prime\,+}_{cb}\to\Sigma_b^0)\approx3\widetilde\Gamma(\widehat \Xi^{*\,+}_{cb}\to\Sigma_b^0)
&\approx&\frac32\widetilde\Gamma(\widehat \Xi^{+}_{cb}\to\Sigma^{*\,0}_b)\approx
\frac12\widetilde\Gamma(\widehat \Xi^{\prime\,+}_{cb}\to\Sigma^{*\,0}_b)\nonumber\\
0.0930\approx0.120&\approx&0.0946\approx0.0813,
\label{eq:rel4}\\\nonumber\\
\widetilde\Gamma(\widehat \Omega^{\prime\,0}_{cb}\to\Xi^{\prime\,-}_b)\approx3
\widetilde\Gamma(\widehat \Omega^{*\,0}_{cb}\to\Xi^{\prime\,-}_b)
&\approx&\frac32\widetilde\Gamma(\widehat \Omega^{0}_{cb}\to\Xi^{*\,-}_b)\approx
\frac12\widetilde\Gamma(\widehat \Omega^{\prime\,0}_{cb}\to\Xi^{*\,-}_b)\nonumber\\
0.0776\approx0.101&\approx&0.0826\approx0.0714,
\label{eq:rel5}\\\nonumber\\
\widetilde\Gamma(\widehat \Xi^{\prime\,+}_{cb}\to\Xi_b^{\prime\,0})\approx3
\widetilde\Gamma(\widehat \Xi^{*\,+}_{cb}\to\Xi_b^{\prime\,0})
&\approx&\frac32\widetilde\Gamma(\widehat \Xi^{+}_{cb}\to\Xi^{*\,0}_b)\approx
\frac12\widetilde\Gamma(\widehat \Xi^{\prime\,+}_{cb}\to\Xi^{*\,0}_b)\nonumber\\
1.65\approx2.24&\approx&1.74\approx1.47,
\label{eq:rel6}\\\nonumber\\
\widetilde\Gamma(\widehat \Omega^{\prime\,0}_{cb}\to\Omega^{-}_b)\approx3
\widetilde\Gamma(\widehat \Omega^{*\,0}_{cb}\to\Omega^{-}_b)
&\approx&\frac32\widetilde\Gamma(\widehat \Omega^{0}_{cb}\to\Omega^{*\,-}_b)\approx
\frac12\widetilde\Gamma(\widehat \Omega^{\prime\,0}_{cb}\to\Omega^{*\,-}_b)\nonumber\\
2.98\approx4.05&\approx&3.57\approx3.01,
\label{eq:rel7}
\eea
\bea
\widetilde\Gamma(\widehat \Xi^{*\,+}_{cb}\to\Sigma^{*\,0}_b)&\approx&
\widetilde\Gamma(\widehat \Xi^{*\,+}_{cb}\to\Sigma_b^0)+\widetilde\Gamma(\widehat \Xi_{cb}^{+}\to\Sigma_b^0)\nonumber\\
0.246&\approx&0.258,
\label{eq:rel8}\\\nonumber\\
\widetilde\Gamma(\widehat \Omega^{*\,0}_{cb}\to\Xi^{*\,-}_b)&\approx&
\widetilde\Gamma(\widehat \Omega^{*\,0}_{cb}\to\Xi'^-_b)+\widetilde\Gamma(\widehat \Omega^{0}_{cb}\to\Xi'^-_b)\nonumber\\
0.223&\approx&0.213,
\label{eq:rel9}\\\nonumber\\
\widetilde\Gamma(\widehat \Xi^{*\,+}_{cb}\to\Xi^{*\,0}_b)&\approx&
\widetilde\Gamma(\widehat \Xi^{*\,+}_{csb}\to\Xi'^0_b)+\widetilde\Gamma(\widehat \Xi^{+}_{cb}\to\Xi'^0_b)\nonumber\\
5.03&\approx&4.99,\label{eq:rel10}\\\nonumber\\
\widetilde\Gamma(\widehat \Omega^{*\,0}_{cb}\to\Omega^{*\,-}_b)&\approx&
\widetilde\Gamma(\widehat \Omega^{*\,0}_{cb}\to\Omega^-_b)+\widetilde\Gamma(\widehat \Omega^{0}_{cb}\to\Omega^-_b)\nonumber\\
10.2&\approx&9.16.
\label{eq:rel11}
\eea
We find our results agree in most of the cases at 
the level of 10\% with some notable
exceptions  in Eqs.~(\ref{eq:rel4}), (\ref{eq:rel5}), (\ref{eq:rel6})
and (\ref{eq:rel7}). These latter discrepancies are largely due, not to the 
the use of the approximate HQSS inspired relations in Eqs.(\ref{eq:aprox1})
-(\ref{eq:aprox9}), but to the fact that the
different baryons that appear in the relations do not have the same
mass, and therefore the available phase space is different for each
transition.  For instance if we just 
make the masses of  ${\widehat\Xi^*_{cb}},{\widehat\Xi^{}_{cb}}$ equal
to the  ${\widehat\Xi^{\prime}_{cb}}$ mass and the 
mass of $\Xi^*_b$ equal to the $\Xi'_b$ mass we
get
\bea
\widetilde\Gamma(\Xi^{\prime\,+}_{cb}\to\Xi_b^{\prime\,0})\approx3
\widetilde\Gamma(\Xi^{*\,+}_{cb}\to\Xi_b^{\prime\,0})
&\approx&\frac32\widetilde\Gamma(\Xi^{+}_{cb}\to\Xi^{*\,0}_b)\approx
\frac12\widetilde\Gamma(\Xi^{\prime\,+}_{cb}\to\Xi^{*\,0}_b)\nonumber\\
1.65\approx1.69&\approx&1.66\approx1.65,
\eea
or in the $\Omega$ sector, with similar changes in the masses,
\bea
\widetilde\Gamma(\Omega^{\prime\,+}_{cb}\to\Omega_b^{\prime\,0})\approx3
\widetilde\Gamma(\Omega^{*\,+}_{cb}\to\Omega_b^{\prime\,0})
&\approx&\frac32\widetilde\Gamma(\Omega^{+}_{cb}\to\Omega^{*\,0}_b)\approx
\frac12\widetilde\Gamma(\Omega^{\prime\,+}_{cb}\to\Omega^{*\,0}_b)\nonumber\\
2.98\approx3.07&\approx&2.93\approx2.91.
\eea
The agreement improves considerably. Then, the HQSS derived relations
are appropriate to evaluate the hadronic amplitudes but  the final results 
may be very sensitive on actual mass values.

Thus, mass differences and the variations induced by them in the
available phase space can not be neglected. Besides the physical
states $B^{(1)}_{cb},B^{(2)}_{cb}$ are not exactly equal to the
$\widehat B_{cb},\widehat B^{\prime}_{cb}$ states and this could also
affect some of the decay widths. In what follows we give the
corresponding numbers for the physical states.
\bea
\widetilde\Gamma( \Xi^{(1)\,+}_{cb}\to\Lambda_b^0)&\approx&\widetilde\Gamma( \Xi^{*\,+}_{cb}\to\Lambda_b^0)
\nonumber\\0.219&\approx&0.235,
\\\nonumber\\
\widetilde\Gamma(\Omega^{(1)\,0}_{cb}\to\Xi^-_b)&\approx&\widetilde\Gamma(\Omega^{*\,0}_{cb}\to\Xi^-_b)
\nonumber\\  0.179&\approx& 0.196,
\\\nonumber\\
\widetilde\Gamma(\Xi^{(1)\,+}_{cb}\to\Xi^0_b)&\approx&\widetilde\Gamma(\Xi^{*\,+}_{cb}\to\Xi^0_b)
\nonumber\\   3.73&\approx& 4.08,
\eea
\bea
\widetilde\Gamma(\Xi^{(2)\,+}_{cb}\to\Sigma_b^0)\approx3\widetilde\Gamma(\Xi^{*\,+}_{cb}\to\Sigma_b^0)
&\approx&\frac32\widetilde\Gamma(\Xi^{(1)\,+}_{cb}\to\Sigma^{*\,0}_b)\approx
\frac12\widetilde\Gamma(\Xi^{(2)\,+}_{cb}\to\Sigma^{*\,0}_b)\nonumber\\
0.110\approx0.120&\approx&0.121\approx0.0737,
\\\nonumber\\
\widetilde\Gamma(\Omega^{(2)\,0}_{cb}\to\Xi^{\prime\,-}_b)\approx3
\widetilde\Gamma(\Omega^{*\,0}_{cb}\to\Xi^{\prime\,-}_b)
&\approx&\frac32\widetilde\Gamma(\Omega^{(1)\,0}_{cb}\to\Xi^{*\,-}_b)\approx
\frac12\widetilde\Gamma(\Omega^{(2)\,0}_{cb}\to\Xi^{*\,-}_b)\nonumber\\
0.0907\approx0.101&\approx&0.104\approx0.0652,
\\\nonumber\\
\widetilde\Gamma(\Xi^{(2)\,+}_{cb}\to\Xi_b^{\prime\,0})\approx3
\widetilde\Gamma(\Xi^{*\,+}_{cb}\to\Xi_b^{\prime\,0})
&\approx&\frac32\widetilde\Gamma(\Xi^{(1)\,+}_{cb}\to\Xi^{*\,0}_b)\approx
\frac12\widetilde\Gamma(\Xi^{(2)\,+}_{cb}\to\Xi^{*\,0}_b)\nonumber\\
1.95\approx2.24&\approx&2.29\approx1.34,
\\\nonumber\\
\widetilde\Gamma(\Omega^{(2)\,0}_{cb}\to\Omega^{-}_b)\approx3
\widetilde\Gamma(\Omega^{*\,0}_{cb}\to\Omega^{-}_b)
&\approx&\frac32\widetilde\Gamma(\Omega^{(1)\,0}_{cb}\to\Omega^{*\,-}_b)\approx
\frac12\widetilde\Gamma(\Omega^{(2)\,0}_{cb}\to\Omega^{*\,-}_b)\nonumber\\
3.49\approx4.05&\approx&4.48\approx2.75,
\eea
\bea
\widetilde\Gamma(\Xi^{*\,+}_{cb}\to\Sigma^{*\,0}_b)&\approx&
\widetilde\Gamma(\Xi^{*\,+}_{cb}\to\Sigma_b^0)+\widetilde\Gamma(\Xi_{cb}^{(1)\,+}\to\Sigma_b^0)\nonumber\\
0.246&\approx&0.238,
\\\nonumber\\
\widetilde\Gamma(\Omega^{*\,0}_{cb}\to\Xi^{*\,-}_b)&\approx&
\widetilde\Gamma(\Omega^{*\,0}_{cb}\to\Xi'^-_b)+\widetilde\Gamma(\Omega^{(1)\,0}_{cb}\to\Xi'^-_b)\nonumber\\
0.223&\approx&0.203,
\\\nonumber\\
\widetilde\Gamma(\Xi^{*\,+}_{cb}\to\Xi^{*\,0}_b)&\approx&
\widetilde\Gamma(\Xi^{*\,+}_{csb}\to\Xi'^0_b)+\widetilde\Gamma(\Xi^{(1)\,+}_{cb}\to\Xi'^0_b)\nonumber\\
5.03&\approx&4.62\\\nonumber\\
\widetilde\Gamma(\Omega^{*\,0}_{cb}\to\Omega^{*\,-}_b)&\approx&
\widetilde\Gamma(\Omega^{*\,0}_{cb}\to\Omega^-_b)+\widetilde\Gamma(\Omega^{(1)\,0}_{cb}\to\Omega^-_b)\nonumber\\
10.2&\approx&8.56.
\eea
Most of the relations are satisfied
 at the 10\% level with a few notable exceptions that involve the 
 decay widths for the $\Xi^{(2)}_{cb}\to\Sigma^{*}_b,\,\Xi^*_b$ 
 and $\Omega^{(2)\,0}_{cb}\to\Xi^{*\,-}_b,\,\Omega^{*\,-}_b$ transitions.

\section{Summary}
\label{sec:summa}
We have made a systematic study of semileptonic decays of $cb$
ground-state doubly heavy baryons driven by $c\to s,d$ transitions at
the quark level. We have employed a simple constituent quark model
scheme, which benefits from the important simplifications in the solution of
 the non-relativistic three body problem that stem from the application of
HQSS~\cite{Albertus:2003sx,Albertus:2006wb}. Despite the modulus of
CKM matrix elements $|V_{cs}|,\,|V_{cd}|$ are much larger than
$|V_{cb}|$, the smaller available phase space leads to 
 $c\to s$ decay widths that turn
out to be larger but of the same order of magnitude as the $b\to c$
driven processes, while widths for $c\to d$ transitions are much smaller.

 As for $b\to c$ semileptonic~\cite{pervin2,Albertus:2009ww} and
 electromagnetic~\cite{Albertus:2010hi,Branz:2010pq} decays, here also
 hyperfine mixing effects have a tremendous impact on $c\to s,d$
 semileptonic decays of spin-1/2 $cb$ baryons. We find factors of 2
 corrections in many cases due to mixing. 

We have derived for the first time HQSS relations for the hadronic
amplitudes.  By requiring invariance under separate bottom and charm
quark spin rotations, we have obtained constraints on the form factors
that enormously simplify the description of these decays. Though,
these relations are strictly valid in the limit of very large heavy
quark masses and near zero recoil, they turn out to be reasonable
accurate for the whole available phase space in these decays. Indeed,
we find our calculation is consistent with HQSS and only deviations at the
10\% level are observed due to the actual, finite, heavy quark
masses. With the use of the HQSS relations and assuming $M_{B_{cb}}=
M_{B'_{cb}} = M_{B^*_{cb}}$ and $M_{B_{b}}= M_{B'_{b}}= M_{B^*_{b}}$,
we have made model independent, though approximate, predictions for
ratios of decay widths. Our values for those ratios agree with the
HQSS motivated predictions at the level of 10\% in most of the
cases. We expect those predictions to hold to that level of accuracy
in other approaches.

\begin{acknowledgments}
  This research was supported by DGI and FEDER funds, under contracts
   FIS2011-28853-C02-02, FPA2010-21750-C02-02, and the Spanish
  Consolider-Ingenio 2010 Programme CPAN (CSD2007-00042),  by Generalitat
  Valenciana under contract PROMETEO/20090090 and by the EU
  HadronPhysics2 project, grant agreement no. 227431. C. A. thanks a Juan de 
  la Cierva contract from the
Spanish  Ministerio de Educaci\'on y Ciencia.
\end{acknowledgments}

\appendix
\section{Nonrelativistic baryon states and wave functions}
\label{app:nrbs}
We construct our nonrelativistic states as follows
\begin{eqnarray}
\label{eq:wf}
&&\hspace{-1cm}\big|{B,r\,\vec{P}}\,\big\rangle_{NR}
=\sqrt{2E}\int d^{\,3}Q_1 \int d^{\,3}Q_2\ {\cal S}_B \sum_{\alpha_1,\alpha_2,\alpha_3}
\widehat{\psi}^{(B,r)}_{\alpha1\,\alpha_2\,\alpha_3}(\,\vec{Q}_1,\vec{Q}_2\,)
\ \frac{1}{(2\pi)^3\ \sqrt{2E_{f_1}2E_{f_2}
2E_{f_3}}}\nonumber\\ 
&&\hspace{3cm}
\times\big|\ \alpha_1\
\vec{p}_1=\frac{m_{f_1}}{\overline{M}}\vec{P}+\vec{Q}_1\ \big\rangle
\big|\ \alpha_2\ \vec{p}_2=\frac{m_{f_2}}{\overline{M}}\vec{P}+\vec{Q}_2\ \big\rangle
\big|\ \alpha_3\ \vec{p}_3=\frac{m_{f_3}}{\overline{M}}\vec{P}-\vec{Q}_1
-\vec{Q}_2\ \big\rangle.
 \end{eqnarray}
The factor $\sqrt{2E}$ is introduced for convenience in order to have
the proper normalization. We denote by $\alpha_j$  the spin ($s$), flavor
($f$) and color ($c$) quantum numbers ( $\alpha\equiv (s,f,c)$\,) of the
$j-$th quark with $(E_{f_j},\,\vec{p}_j)$ and $m_{f_j}$ its
four-momentum and mass, and $\overline{M}=m_{f_1}+m_{f_2}+m_{f_3}$.  
Individual quark states are
normalized such that $\left\langle\ \alpha^{\prime}\ \vec{p}^{\
\prime}\,|\,\alpha\ \vec{p}\, \right\rangle=2E_f\, (2\pi)^3\,
\delta_{\alpha^{\prime}\, \alpha}\,\delta^3( \vec{p}^{\
\prime}-\vec{p}\,)$.  $\widehat{\psi}^{\,(B,r)}_{\alpha_1\,\alpha_2\,\alpha_3}
(\,\vec{Q}_1,\vec{Q}_2\,)$ is the internal wave function in momentum
space, being $\vec{Q}_1$ ($\vec{Q}_2$) the conjugate momenta to the
relative position $\vec{r}_1$ ($\vec{r}_2$) between quark 1 (2) and the
 third quark.  In the transitions under study an initial $c\,b\,l'$
baryon decays into a final $l\,l'\,b$ one, where $l=d,s$ and
$l'=u,d,s$. We construct the wave functions such that the   $c$ and $b$
quarks in the initial baryon are quarks 1 and 2 respectively. Also in the 
final baryon  the two light quarks $l$ and $l'$ are respectively quarks 
1 and 2. 
Expressions for the different
$\widehat{\psi}^{(B,r)}_{\alpha_1\,\alpha_2\,\alpha_3}
(\,\vec{Q}_1,\vec{Q}_2\,)$ are given  below.
These wave functions are normalized as
\begin{equation}
\int d^{\,3}Q_1 \int d^{\,3}Q_2\ \sum_{\alpha_1,\alpha_2,\alpha_3}
\left(\widehat{\psi}^{(B,r')}_{\alpha_1\,\alpha_2\,\alpha_3}(\,\vec{Q}_1,\vec{Q}_2\,)\right)^*
\widehat{\psi}^{(B,r)}_{\alpha_1\,\alpha_2\,\alpha_3}(\,\vec{Q}_1,\vec{Q}_2\,)
=\delta_{rr'}.
\end{equation}
For the final states we use wave functions that are antisymmetric
under the exchange of quarks 1 and 2 quantum numbers. In order for our 
nonrelativistic baryon states to have the proper normalization
\bea
{}_{\stackrel{}{NR}}
\big\langle\, {B,r'\,\vec{P}^{\,\prime}}\,|\,{B,r
\,\vec{P}}\,\big\rangle_{NR}
=2E\,(2\pi)^3\,\delta_{rr'}\,\delta^3(\vec{P}^{\,\prime}-\vec{P}\,).
\eea
 we need
 to introduce in
Eq.~(\ref{eq:wf}) a symmetry factor ${\cal S}_B=\frac1{\sqrt2}$ for those 
states. For the initial states  ${\cal S}_B=1$.

The wave functions for $cb$ states where the spin of the heavy quark subsystem
is well defined are given by
\begin{eqnarray}
\widehat{\psi}^{\,(\Xi^{+}_{cb},s)}_{\alpha_1\,\alpha_2\,\alpha_3}(\,\vec{Q}_1,\vec{Q}_2\,)
&=&\frac{1}{\sqrt{3!}}\,\varepsilon_{c_1\,c_2\,c_3}\
\widehat{\phi}^{\,(\Xi^{+}_{cb}, s)}_{(s_1,\,f_1)\ ,(s_2,\,f_2)\,(s_3,\,f_3)}
(\,\vec{Q}_1, \vec{Q}_2\,)\nonumber\\
&=&\frac{1}{\sqrt{3!}}\,\varepsilon_{c_1\,c_2\,c_3}\
\ \widetilde{\phi}^{\,(\Xi_{cb})}(\,\vec{Q}_1,\vec{Q}_2\,)\ 
\delta_{f_1\,c}\,\delta_{f_2\,b}\, \delta_{f_3\,u}\nonumber\\
&&\hspace{2cm} \times\  (1/2,1/2,1;s_1,s_2,s_1+s_2)\ 
(1,1/2,1/2;s_1+s_2,s_3,s),
\eea
\bea
\widehat{\psi}^{\,(\Xi^{'+}_{cb},s)}_{\alpha_1\,\alpha_2\,\alpha_3}(\,\vec{Q}_1,\vec{Q}_2\,)
&=&\frac{1}{\sqrt{3!}}\,\varepsilon_{c_1\,c_2\,c_3}\
\widehat{\phi}^{\,(\Xi^{'+}_{cb}, s)}_{(s_1,\,f_1)\ ,(s_2,\,f_2)\,(s_3,\,f_3)}
(\,\vec{Q}_1, \vec{Q}_2\,)\nonumber\\
&=&\frac{1}{\sqrt{3!}}\,\varepsilon_{c_1\,c_2\,c_3}\
\ \widetilde{\phi}^{\,(\Xi_{cb})}(\,\vec{Q}_1,\vec{Q}_2\,)\ 
\delta_{f_1\,c}\,\delta_{f_2\,b}\, \delta_{f_3\,u}\  
(1/2,1/2,0;s_1,s_2,0)\,\delta_{s_3\,s},
\eea
\bea
\widehat{\psi}^{\,(\Xi^{*+}_{cb},s)}_{\alpha_1\,\alpha_2\,\alpha_3}(\,\vec{Q}_1,\vec{Q}_2\,)
&=&\frac{1}{\sqrt{3!}}\,\varepsilon_{c_1\,c_2\,c_3}\
\widehat{\phi}^{\,(\Xi^{*+}_{cb}, s)}_{(s_1,\,f_1)\ ,(s_2,\,f_2)\,(s_3,\,f_3)}
(\,\vec{Q}_1, \vec{Q}_2\,)\nonumber\\
&=&\frac{1}{\sqrt{3!}}\,\varepsilon_{c_1\,c_2\,c_3}\
\ \widetilde{\phi}^{\,(\Xi^{*}_{cb})}(\,\vec{Q}_1,\vec{Q}_2\,)\ 
\delta_{f_1\,c}\,\delta_{f_2\,b}\, \delta_{f_3\,u}\nonumber\\
&&\hspace{2cm} \times\  (1/2,1/2,1;s_1,s_2,s_1+s_2)\ 
(1,1/2,3/2;s_1+s_2,s_3,s),
\eea
where
$\varepsilon_{c_1 c_2 c_3}$ is the totally antisymmetric tensor
with  $\frac{\varepsilon_{c_1 c_2 c_3}}{\sqrt{3!}}$ being the fully
antisymmetric color wave
function. The $(j_1,j_2,j;m_1,m_2,m)$ are SU(2) Clebsch-Gordan 
coefficients.
The different
   $\tilde \phi(\,\vec{Q}_1,\vec{Q}_2\,)$ wave functions  have  
   total orbital angular momentum  0 being invariant under
rotations and thus depending only  on
  $|\vec{Q}_1|$, $|\vec{Q}_2|$ 
and $\vec{Q}_1\cdot\vec{Q}_2$. They are normalized such that
\begin{equation}
\int d^{\,3}Q_1 \int d^{\,3}Q_2\ 
\left|\widetilde{\phi}(\,\vec{Q}_1,\vec{Q}_2\,)\right|^2
=1.
\end{equation}
The corresponding $\Xi^{('*)}$ neutral states are obtained by implementing the
trivial replacement $\delta_{f_3\,u} \to \delta_{f_3\,d}$. Besides,
the $\Omega^{('*)}$ color-spin-flavor-momentum wave-functions 
 are obtained from the cascade ones by
substituting the momentum space
$\widetilde{\phi}^{\,\Xi^{('*)}_{cb}}(\,\vec{Q}_1,\vec{Q}_2\,)$ wave
functions by the appropriated
$\widetilde{\phi}^{\,\Omega^{('*)}_{cb}}(\,\vec{Q}_1,\vec{Q}_2\,)$
ones, and always using $\delta_{f_3\,s}$.
For  $b$-heavy baryons  we further have  
\begin{eqnarray}
\widehat{\psi}^{\,(\Lambda^{0}_{b},s)}_{\alpha_1\,\alpha_2\,\alpha_3}(\,\vec{Q}_1,\vec{Q}_2\,)
&=&\frac{1}{\sqrt{3!}}\,\varepsilon_{c_1\,c_2\,c_3}\
\widehat{\phi}^{\,(\Lambda^{0}_{b}, s)}_{(s_1,\,f_1)\ ,(s_2,\,f_2)\,(s_3,\,f_3)}
(\,\vec{Q}_1, \vec{Q}_2\,)\nonumber\\
&=&\frac{1}{\sqrt{3!}}\,\varepsilon_{c_1\,c_2\,c_3}\
\ \widetilde{\phi}^{\,(\Lambda^{0}_{b})}(\,\vec{Q}_1,\vec{Q}_2\,)\ 
\frac{1}{\sqrt2}(\delta_{f_1\,u}\, \delta_{f_2\,d}-
\delta_{f_1\,d}\, \delta_{f_2\,u})\delta_{f_3\,b}
(1/2,1/2,0;s_1,s_2,0)\,\delta_{s_3\,s}, 
\eea
\bea
\widehat{\psi}^{\,(\Sigma^{0}_{b},s)}_{\alpha_1\,\alpha_2\,\alpha_3}(\,\vec{Q}_1,\vec{Q}_2\,)
&=&\frac{1}{\sqrt{3!}}\,\varepsilon_{c_1\,c_2\,c_3}\
\widehat{\phi}^{\,(\Sigma^{0}_{b}, s)}_{(s_1,\,f_1)\ ,(s_2,\,f_2)\,(s_3,\,f_3)}
(\,\vec{Q}_1, \vec{Q}_2\,)\nonumber\\
&=&\frac{1}{\sqrt{3!}}\,\varepsilon_{c_1\,c_2\,c_3}\
\ \widetilde{\phi}^{\,(\Sigma_{b})}(\,\vec{Q}_1,\vec{Q}_2\,)\ 
\frac{1}{\sqrt2}(\delta_{f_1\,u}\, \delta_{f_2\,d}+
\delta_{f_1\,d}\, \delta_{f_2\,u})\delta_{f_3\,b}\nonumber\\
&&\hspace{.5cm}\times\,(1/2,1/2,1;s_1,s_2,s_1+s_2)(1,1/2,1/2,s_1+s_2,s_3,s), 
\eea
\bea
\widehat{\psi}^{\,(\Sigma^{*0}_{b},s)}_{\alpha_1\,\alpha_2\,\alpha_3}(\,\vec{Q}_1,\vec{Q}_2\,)
&=&\frac{1}{\sqrt{3!}}\,\varepsilon_{c_1\,c_2\,c_3}\
\widehat{\phi}^{\,(\Sigma^{*0}_{b}, s)}_{(s_1,\,f_1)\ ,(s_2,\,f_2)\,(s_3,\,f_3)}
(\,\vec{Q}_1, \vec{Q}_2\,)\nonumber\\
&=&\frac{1}{\sqrt{3!}}\,\varepsilon_{c_1\,c_2\,c_3}\
\ \widetilde{\phi}^{\,(\Sigma^{*}_{b})}(\,\vec{Q}_1,\vec{Q}_2\,)\ 
\frac{1}{\sqrt2}(\delta_{f_1\,u}\, \delta_{f_2\,d}+
\delta_{f_1\,d}\, \delta_{f_2\,u})\delta_{f_3\,b}\nonumber\\
&&\hspace{.5cm}\times\,(1/2,1/2,1;s_1,s_2,s_1+s_2)(1,1/2,3/2,s_1+s_2,s_3,s),
\eea
\bea
\widehat{\psi}^{\,(\Xi^0_{b},s)}_{\alpha_1\,\alpha_2\,\alpha_3}(\,\vec{Q}_1,\vec{Q}_2\,)
&=&\frac{1}{\sqrt{3!}}\,\varepsilon_{c_1\,c_2\,c_3}\
\widehat{\phi}^{\,(\Xi^0_{b}, s)}_{(s_1,\,f_1)\,(s_2,\,f_2)\,(s_3,\,f_3)}
(\,\vec{Q}_1, \vec{Q}_2\,)\nonumber\\
&=&\frac{1}{\sqrt{3!}}\,\varepsilon_{c_1\,c_2\,c_3}\
\ \frac1{\sqrt2}\,(\widetilde{\phi}^{\,(\Xi^0_{b})}_{us}(\,\vec{Q}_1,\vec{Q}_2\,)\ 
\delta_{f_1\,u}\, \delta_{f_2\,s}-\widetilde{\phi}^{\,(\Xi^0_{b})}_{su}(\,\vec{Q}_1,\vec{Q}_2\,)\ 
\delta_{f_1\,s}\, \delta_{f_2\,u}
)\, \delta_{f_3\,b}\nonumber\\
&&\hspace{2cm} \times\   (1/2,1/2,0;s_1,s_2,0)\,\delta_{s_3\,s}, 
\eea
\bea
\widehat{\psi}^{\,(\Xi'^{\,0}_{b},s)}_{\alpha_1\,\alpha_2\,\alpha_3}(\,\vec{Q}_1,\vec{Q}_2\,)
&=&\frac{1}{\sqrt{3!}}\,\varepsilon_{c_1\,c_2\,c_3}\
\widehat{\phi}^{\,(\Xi'^{\, 0}_{b}, s)}_{(s_1,\,f_1)\,(s_2,\,f_2)\,(s_3,\,f_3)}
(\,\vec{Q}_1, \vec{Q}_2\,)\nonumber\\
&=&\frac{1}{\sqrt{3!}}\,\varepsilon_{c_1\,c_2\,c_3}\
\ \frac1{\sqrt2}\,(\widetilde{\phi}^{\,(\Xi'^{\,0}_{b})}_{us}(\,\vec{Q}_1,\vec{Q}_2\,)\ 
\delta_{f_1\,u}\, \delta_{f_2\,s}+\widetilde{\phi}^{\,(\Xi'^{\,0}_{b})}_{su}(\,\vec{Q}_1,\vec{Q}_2\,)\ 
\delta_{f_1\,s}\, \delta_{f_2\,u}
)\, \delta_{f_3\,b}\nonumber\\
&&\hspace{2cm} \times\   (1/2,1/2,1;s_1,s_2,s_1+s_2)\ 
(1,1/2,1/2;s_1+s_2,s_3,s), 
\eea
\bea
\widehat{\psi}^{\,(\Xi^{*\, 0}_{b},s)}_{\alpha_1\,\alpha_2\,\alpha_3}(\,\vec{Q}_1,\vec{Q}_2\,)
&=&\frac{1}{\sqrt{3!}}\,\varepsilon_{c_1\,c_2\,c_3}\
\widehat{\phi}^{\,(\Xi^{*\, 0}_{b}, s)}_{(s_1,\,f_1)\,(s_2,\,f_2)\,(s_3,\,f_3)}
(\,\vec{Q}_1, \vec{Q}_2\,)\nonumber\\
&=&\frac{1}{\sqrt{3!}}\,\varepsilon_{c_1\,c_2\,c_3}\
\ \frac1{\sqrt2}\,(\widetilde{\phi}^{\,(\Xi^{*\, 0}_{b})}_{us}(\,\vec{Q}_1,\vec{Q}_2\,)\ 
\delta_{f_1\,u}\, \delta_{f_2\,s}
+\widetilde{\phi}^{\,(\Xi^{*\, 0}_{b})}_{su}(\,\vec{Q}_1,\vec{Q}_2\,)\ 
\delta_{f_1\,s}\, \delta_{f_2\,u})\, \delta_{f_3\,b}\nonumber\\
&&\hspace{2cm} \times\   (1/2,1/2,1;s_1,s_2,s_1+s_2)\ 
(1,1/2,3/2;s_1+s_2,s_3,s), 
\eea
\bea
\widehat{\psi}^{\,(\Omega^{-}_{b},s)}_{\alpha_1\,\alpha_2\,\alpha_3}(\,\vec{Q}_1,\vec{Q}_2\,)
&=&\frac{1}{\sqrt{3!}}\,\varepsilon_{c_1\,c_2\,c_3}\
\widehat{\phi}^{\,(\Omega^{-}_{b}, s)}_{(s_1,\,f_1)\ ,(s_2,\,f_2)\,(s_3,\,f_3)}
(\,\vec{Q}_1, \vec{Q}_2\,)\nonumber\\
&=&\frac{1}{\sqrt{3!}}\,\varepsilon_{c_1\,c_2\,c_3}\
\ \widetilde{\phi}^{\,(\Omega^{-}_{b})}(\,\vec{Q}_1,\vec{Q}_2\,)\ 
\delta_{f_1\,s}\, \delta_{f_2\,s}\,\delta_{f_3\,b}\nonumber\\
&&\hspace{.5cm}\times\,(1/2,1/2,1;s_1,s_2,s_1+s_2)(1,1/2,1/2,s_1+s_2,s_3,s) 
\eea
\bea
\widehat{\psi}^{\,(\Omega^{*-}_{b},s)}_{\alpha_1\,\alpha_2\,\alpha_3}(\,\vec{Q}_1,\vec{Q}_2\,)
&=&\frac{1}{\sqrt{3!}}\,\varepsilon_{c_1\,c_2\,c_3}\
\widehat{\phi}^{\,(\Omega^{*-}_{b}, s)}_{(s_1,\,f_1)\ ,(s_2,\,f_2)\,(s_3,\,f_3)}
(\,\vec{Q}_1, \vec{Q}_2\,)\nonumber\\
&=&\frac{1}{\sqrt{3!}}\,\varepsilon_{c_1\,c_2\,c_3}\
\ \widetilde{\phi}^{\,(\Omega^{*-}_{b})}(\,\vec{Q}_1,\vec{Q}_2\,)\ 
\delta_{f_1\,s}\, \delta_{f_2\,s}\,\delta_{f_3\,b}\nonumber\\
&&\hspace{.5cm}\times\,(1/2,1/2,1;s_1,s_2,s_1+s_2)(1,1/2,3/2,s_1+s_2,s_3,s). 
\eea
Here, besides the properties above, the relation
$\widetilde{\phi}_{sn}(\,\vec{Q}_1,\vec{Q}_2\,)=
\widetilde{\phi}_{ns}(\,\vec{Q}_2,\vec{Q}_1\,)$, with $n=u,d$, also
applies. The wave functions for the other  members of the different
isospin multiplets are obtained from those given above by implementing
obvious substitutions.

The momentum space wave functions are the Fourier transform of the
corresponding wave functions in coordinate space,
\be
\widetilde{\phi}(\,\vec{Q}_1,\vec{Q}_2\,) = \frac{1}{(2\pi)^3} \int
d^3 r_1 d^3 r_2 e^{-i \vec{Q}_1\cdot \vec{r}_1}e^{-i \vec{Q}_2\cdot
  \vec{r}_2} \phi (\,\vec{r}_1,\vec{r}_2\,).
\ee
We use a HQSS constrained variational approach to deal with the
underlying three body problem and to obtain the spatial wave
functions. For the latter we consider they only  depend on the
three interquark relative distances $r_1$, $r_2$ and $r_{12}=|\,
\vec{r}_1-\vec{r}_2|$. This amounts to assume that the total orbital
angular momentum of the baryon is zero. However, this does not imply
that the individual orbital angular momenta ($l_{13}$ and $l_{23}$ )
of the $(13)$ and $(23)$ pairs is zero, though both $l_{13}$ and
$l_{23}$ should take a common value $l$, since $\vec l_{13}$ and $\vec
l_{23}$ must be coupled to a total $S$-wave. Indeed, the wave
functions $\phi (\,\vec{r}_1,\vec{r}_2\,)$ can be decomposed as a sum
of a large number of contributions or multipoles for different values
of $l=0,1,2,3...$. More details, for the case of singly and doubly
heavy baryons can be found in
Refs.~\cite{Albertus:2003sx,Albertus:2006wb}, respectively.

As already mentioned the two baryons states $\Xi_{cb},\,\Xi'_{cb}$
differ just in the spin of the heavy degrees of freedom, and thus they
mix under the effect of the hyperfine interaction between the light
quark and any of the heavy quarks. The same happens for the
$\Omega_{cb},\,\Omega'_{cb}$ states. This mixing is important
and greatly affects the results for the decay widths. The mixing is
however negligible for the $\Xi_b,\,\Xi'_b$ and
$\Omega_b,\,\Omega'_b$ states and we have ignored it.

\section{Weak matrix elements and form factors}
\label{app:ffwme}
Taking the initial baryon at rest and $\vec q$ in the positive $Z$ direction 
we define vector and axial matrix elements
\bea
V^\mu_{r\to r'}-A^\mu_{r\to r'}&=&\big\langle B', r'\ \vec{P}^{\,\prime}=-\vec q\left|\,
\overline \Psi_l(0)\gamma^\mu(1-\gamma_5)\Psi_c(0)
 \right| B, r\ \vec{P}=\vec 0
\big\rangle,
\eea
 that  in our model are given as
\bea
V^\mu_{r\to r'}-A^\mu_{r\to r'}&=&\sqrt{2M}\sqrt{2E'}\int d^3Q_1\int d^3Q_2\ 
 \left(\tilde\phi^{(
 B')}(\vec Q_1-\frac{m_b+m_{l'}}{\overline{M'}}\,\vec q,-\vec Q_1-\vec
 Q_2+\frac{m_{l'}}{\overline{M'}}\,\vec q\,)\right)^*\tilde\phi^{(B)}(\vec Q_1,\vec Q_2)\nonumber\\
&&{\cal F}\ \sum_{s_1,s_2}(1/2,1/2,S';r'-r+s_1,r-s_1-s_2,r'-s_2)(S',1/2,J';r'-s_2,s_2,r')
\nonumber\\&&\hspace{1.125cm}(1/2,1/2,S;s_1,s_2,s_1+s_2)
(S,1/2,J;s_1+s_2,r-s_1-s_2,r)\nonumber\\
&&\hspace{2cm} \frac{\overline{u}_{l\
r'-r+s_1}(\vec{Q}_1-\vec{q}\,)\ \gamma^\mu(1-\gamma_5)\ u_{c\,
s_1}(\vec{Q}_1)}{\sqrt{2E_l(|\vec{Q}_1-\vec{q}\,|)2E_c(|\vec{Q}_1|)}},
\label{eq:vame}
\eea
where $J,S$ ($J',S'$) are the total spin and the spin of the two first quarks for
the initial (final ) baryon. ${\cal F}$ is a flavor factor that depends on the
transitions and which values are collected in Table~\ref{tab:ffactors}.
Here we have a $c\to l$ transition at the quark level, while $l'$ is
the light quark originally present in the initial baryon. When the final baryon has just one $s$ quark
then $\tilde\phi^{( B')}$ should be interpreted respectively as $\tilde\phi^{(
B')}_{sn}$ or $\tilde\phi^{( B')}_{ds}$  for the case of 
 $c\to s$ or $c\to d$ transitions.

\begin{table}
\begin{tabular}{lc}\hline
&\ \ ${\cal F}$\\\hline\\
$\Xi_{cb}^{+}\to\Xi^0_b$
 &$1$\\
 
$\Xi_{cb}^{0}\to\Xi^-_b$
 &$1$\\

$\Xi_{cb}^{+}\to\Xi^{\prime\,0}_b$
 &$-1$\\

$\Xi_{cb}^{0}\to\Xi^{\prime\,-}_b$
 &$-1$\\

$\Xi_{cb}^{+}\to\Xi^{*\,0}_b$
 &$-1$\\

$\Xi_{cb}^{0}\to\Xi^{*\,-}_b$
 &$-1$\\

$\Xi_{cb}^{\prime\,+}\to\Xi^0_b$
 &$1$\\
 
$\Xi_{cb}^{\prime\,0}\to\Xi^-_b$
 &$1$\\

$\Xi_{cb}^{\prime\,+}\to\Xi^{\prime\,0}_b$
 &$-1$\\
 
$\Xi_{cb}^{\prime\,0}\to\Xi^{\prime\,-}_b$
 &$-1$\\
 
$\Xi_{cb}^{\prime\,+}\to\Xi^{*\,0}_b$
 &$-1$\\

$\Xi_{cb}^{\prime\,0}\to\Xi^{*\,-}_b$
 &$-1$\\

$\Xi_{cb}^{*\,+}\to\Xi^{0}_b$
 &$1$\\

$\Xi_{cb}^{*\,0}\to\Xi^{-}_b$
 &$1$\\

$\Xi_{cb}^{*\,+}\to\Xi^{\prime\,0}_b$
 &$-1$\\

$\Xi_{cb}^{*\,0}\to\Xi^{\prime\,-}_b$
 &$-1$\\

$\Xi^{*\,+}_{cb}\to\Xi^{*\,0}_b$
 &$-1$\\

$\Xi^{*\,0}_{cb}\to\Xi^{*\,-}_b$
 &$-1$\\

$\Omega^{0}_{cb}\to\Omega^{-}_b$
 &$-\sqrt2$\\

$\Omega^{0}_{cb}\to\Omega^{*\,-}_b$
 &$-\sqrt2$\\

$\Omega^{\prime\,0}_{cb}\to\Omega^{-}_b$
 &$-\sqrt2$\\

$\Omega^{\prime\,0}_{cb}\to\Omega^{*\,-}_b$
 &$-\sqrt2$\\
 
$\Omega^{*\,0}_{cb}\to\Omega^{-}_b$
 &$-\sqrt2$\\

$\Omega^{*\,0}_{cb}\to\Omega^{*\,-}_b$
 &$-\sqrt2$\\
\\
 
\hline
\end{tabular}
%
%
%
%
\hspace{.5cm}
\begin{tabular}{lc}\hline
&\ \ ${\cal F}$\vspace{.1cm}\\\hline\\
$\Xi_{cb}^{+}\to\Lambda^0_b$
 &1\\
 
$\Xi_{cb}^{+}\to\Sigma^0_b$
 &$-1$\\

$\Xi_{cb}^{0}\to\Sigma^-_b$
 &$-\sqrt2$\\

$\Xi_{cb}^{+}\to\Sigma^{*\,0}_b$
 &$-1$\\

$\Xi_{cb}^{0}\to\Sigma^{*\,-}_b$
 &$-\sqrt2$\\

$\Xi^{\prime\,+}_{cb}\to\Lambda^0_b$
 &$1$\\
 
$\Xi^{\prime\,+}_{cb}\to\Sigma^0_b$
 &$-1$\\

$\Xi^{\prime\,0}_{cb}\to\Sigma^-_b$
 &$-\sqrt2$\\

$\Xi^{\prime\,+}_{cb}\to\Sigma^{*\,0}_b$
 &$-1$\\

$\Xi^{\prime\,0}_{cb}\to\Sigma^{*\,-}_b$
 &$-\sqrt2$\\
 
$\Xi^{*\,+}_{cb}\to\Lambda^{0}_b$
 &$1$\\
 
$\Xi^{*\,+}_{cb}\to\Sigma^{0}_b$
 &$-1$\\
 
$\Xi^{*\,0}_{cb}\to\Sigma^{-}_b$
 &$-\sqrt2$\\
 
$\Xi^{*\,+}_{cb}\to\Sigma^{*\,0}_b$
 &$-1$\\

$\Xi^{*\,0}_{cb}\to\Sigma^{*\,-}_b$
 &$-\sqrt2$\\

$\Omega^{0}_{cb}\to\Xi^{-}_b$
 &$-1$\\

$\Omega^{0}_{cb}\to\Xi^{\prime\,-}_b$
 &$-1$\\

$\Omega^{0}_{cb}\to\Xi^{*\,-}_b$
 &$-1$\\

$\Omega^{\prime\,0}_{cb}\to\Xi^{-}_b$
 &$-1$\\

$\Omega^{\prime\,0}_{cb}\to\Xi^{\prime\,-}_b$
 &$-1$\\

$\Omega^{\prime\,0}_{cb}\to\Xi^{*\,-}_b$
 &$-1$\\

$\Omega^{*\,0}_{cb}\to\Xi^{-}_b$
 &$-1$\\

$\Omega^{*\,0}_{cb}\to\Xi^{\prime\,-}_b$
 &$-1$\\

$\Omega^{*\,0}_{cb}\to\Xi^{*\,-}_b$
 &$-1$\\\\

\hline
\end{tabular}
\caption{${\cal F}$ flavor factors (Eq.~(\ref{eq:vame}))
for  for $c\to s$ (left panel) and $c\to d$ (right panel) transitions.}
\label{tab:ffactors}
\end{table}

 Relations between different matrix elements  can be
found by performing the spin sums in Eq.~(\ref{eq:vame}). For that purpose the 
following results, that we obtain for $\vec q$ in the positive $Z$ direction, 
are very useful
\bea
\frac{1}{\sqrt{2E_l2E_c}}\bar u_{l\,s'}(\vec p-\vec q\,)
\gamma^0 u_{c\,s}(\vec p\,)&=&
\sqrt\frac{(E_l+m_l)(E_c+m_c)}{2E_l2E_c}\left[ \
\left(1+\frac{\vec p\,^2-|\vec q\,|p^3}{(E_l+m_l)(E_c+m_c)}\right)\,\delta_{s'\,s}
\right.\nonumber\\
&&\hspace{1.cm}+\frac{|\vec
q\,|}{(E_l+m_l)(E_c+m_c)}\left(\ \, (-p^1+ip^2)\,\delta_{s'\,s+1}\right.\left.
+(p^1+ip^2)\,\delta_{s'\,s-1}\right)\bigg],
\eea

\bea
&&\hspace*{-.5cm}\frac{1}{\sqrt{2E_l2E_c}}\bar u_{l\,s'}(\vec p-\vec q\,)
\gamma^j u_{c\,s}(\vec p\,)\nonumber\\
&&=\sqrt\frac{(E_l+m_l)(E_c+m_c)}{2E_l2E_c}\left[\ \ 
\left(\frac{ p^j}{E_c+m_c}+\frac{p^j- q^j}{E_l+m_l}+i\frac{E_c+m_c-E_l-m_l}
{(E_l+m_l)(E_c+m_c)}
(-p^2\delta_{j1}+p^1\delta_{j2})(\delta_{s\,1/2}-
\delta_{s\,-1/2})\right)\,\delta_{s'\,s}\right.\nonumber\\
&&\hspace{3.75cm}+\delta_{j1}\frac{|\vec q\,|(E_c+m_c)-(E_c+m_c-E_l-m_l)p^3}
{(E_l+m_l)(E_c+m_c)}
\left(\delta_{s'\,s-1}-\delta_{s'\,s+1}
\right)\nonumber\\
&&\hspace{3.75cm}\left.+i\delta_{j2}\frac{|\vec q\,|(E_c+m_c)-(E_c+m_c-E_l-m_l)p^3}
{(E_l+m_l)(E_c+m_c)}
\left(\delta_{s'\,s+1}+\delta_{s'\,s-1}
\right)\right.\nonumber\\
&&\hspace{3.75cm}\left.+\delta_{j3}
\frac{E_c+m_c-E_l-m_l}{(E_l+m_l)(E_c+m_c)}((-p^1+ip^2)\delta_{s'\,s+1}+(
p^1+ip^2)\delta_{s'\,s-1})
\right],
\eea

\bea
\frac{1}{\sqrt{2E_l2E_c}}\bar u_{l\,s'}(\vec p-\vec q\,)
\gamma^0\gamma_5 u_{c\,s}(\vec p\,)&=&\sqrt\frac{(E_l+m_l)(E_c+m_c)}{2E_l2E_c}\left[ \ \ 
\left(\frac{p^3}{E_c+m_c}+\frac{p^3-|\vec q\,|}{E_l+m_l}\right)
(\delta_{s\,1/2}-\delta_{s\,-1/2})\,\delta_{s'\,s}
\right.\nonumber\\
&&\hspace{.5cm}+\left(\frac{1}{E_c+m_c}+\frac{1}{E_l+m_l}\right)\left(\ \, (p^1-ip^2)\,\delta_{s'\,s+1}\right.\left.
+(p^1+ip^2)\,\delta_{s'\,s-1}\right)\bigg],
\eea
\bea
&&\hspace*{-.5cm}\frac{1}{\sqrt{2E_l2E_c}}\bar u_{l\,s'}(\vec p-\vec q\,)
\gamma^j\gamma_5 u_{c\,s}(\vec p\,)\nonumber\\
&&=\sqrt\frac{(E_l+m_l)(E_c+m_c)}{2E_l2E_c}\left[\ 
\left(\frac{i(\vec q\times\vec p)^j
}{(E_l+m_l)(E_c+m_c)}\right)\,\delta_{s'\,s}\right.\nonumber\\
&&\hspace{3.75cm}\left.+
\left(1-\frac{\vec p\,^2-|\vec q\,|p^3
}{(E_l+m_l)(E_c+m_c)}\right)\bigg(\
\delta_{s'\,s}\delta_{j3}(\delta_{s\,1/2}-\delta_{s\,-1/2})+
\delta_{s'\,s+1}(\delta_{j1}-i\delta_{j2})\right.\nonumber\\
&&\hspace{9cm}+\delta_{s'\,s-1}(\delta_{j1}+i\delta_{j2})\bigg)\nonumber\\
&&\hspace{3.75cm}+\frac{2p^j-q^j}{(E_l+m_l)(E_c+m_c)}((p^1-ip^2)\delta_{s'\,s+1}
+(p^1+ip^2)\delta_{s'\,s-1})\nonumber\\
&&\hspace{3.75cm}\left.+\frac{(p^j-q^j)p^3+p^j(p^3-|\vec q\,|)}{(E_l+m_l)(E_c+m_c)}\delta_{s'\,s}
(\delta_{s\,1/2}-\delta_{s\,-1/2})\right].
\eea

The fact that the orbital wave functions are invariant under rotations implies
that  the integrals of the form
\bea
&&\hspace{-.5cm}I^j(\vec q\,)=\int d^3Q_1\int d^3Q_2\ 
 \left(\tilde\phi^{(
 B')}(\vec Q_1-\frac{m_b+m_{l'}}{\overline{M'}}\,\vec q,-\vec Q_1-\vec
 Q_2+\frac{m_{l'}}{\overline{M'}}\,\vec q\,)\right)^*\tilde\phi^{(B)}(\vec Q_1,\vec Q_2)
 \ F(|\vec Q_1-\vec q\,|, |Q_1|)\ Q_1^j,\nonumber\\
&&\hspace{-.5cm}I^{jk}(\vec q\,)=\int d^3Q_1\int d^3Q_2\ 
 \left(\tilde\phi^{(
 B')}(\vec Q_1-\frac{m_b+m_{l'}}{\overline{M'}}\,\vec q,-\vec Q_1-\vec
 Q_2+\frac{m_{l'}}{\overline{M'}}\,\vec q\,)\right)^*\tilde\phi^{(B)}(\vec Q_1,\vec Q_2)
 \ F(|\vec Q_1-\vec q\,|, |Q_1|)\ Q_1^j Q_1^k,\nonumber\\
\eea
where $F(|\vec Q_1-\vec q\,|, |\vec Q_1|)$ is a function of $|\vec Q_1-\vec q\,|$ and 
$|\vec Q_1|$, are tensors under rotations and are thus given by
\bea
I^j(\vec q\,)&=&C(|\vec q\,|) \frac{q^j}{|\vec q\,|}, \nonumber\\
I^{jk}(\vec q\,)&=&D(|\vec q\,|)\delta^{jk}+E(|\vec q\,|)
\frac{q^jq^k}{|\vec q\,|^2}.
\eea
As a result we have that $I^1(\vec q\,)=I^2(\vec q\,)=0$,
$I^{11}(\vec q\,)=I^{22}(\vec q\,)$ and 
$I^{jk}(\vec q\,)=0$ unless $j=k$.
 With all this in mind, one can see that all 
 spin sums that appear in the evaluation of the different matrix elements 
 correspond to
one of the following cases
\begin{enumerate}
\item $V^0_{r\to r'},\ V^3_{r\to r'}$
\bea
&&\sum_{s_1,s_2}(1/2,1/2,S';r'-r+s_1,r-s_1-s_2,r'-s_2)(S',1/2,J';r'-s_2,s_2,r')
\nonumber\\&&\hspace{.65cm}(1/2,1/2,S;s_1,s_2,s_1+s_2)
(S,1/2,J;s_1+s_2,r-s_1-s_2,r)\ \delta_{r\,r'}\nonumber\\
&&\hspace{1cm}=\big\langle[(1/2_{(1)}\otimes1/2_{(3)})^{S'}\otimes1/2_{(2)}]^{J'}_{r'}\big|
[(1/2_{(1)}\otimes1/2_{(2)})^S\otimes1/2_{(3)}]^{J}_r\big\rangle,
\nonumber\\
\eea
\item $V^1_{r\to r'},\ V^2_{r\to r'}$
\bea
&&\sum_{s_1,s_2}(1/2,1/2,S';r'-r+s_1,r-s_1-s_2,r'-s_2)(S',1/2,J';r'-s_2,s_2,r')
\nonumber\\&&\hspace{0.65cm}(1/2,1/2,S;s_1,s_2,s_1+s_2)
(S,1/2,J;s_1+s_2,r-s_1-s_2,r)\ \delta_{r'-r\ \pm1}\nonumber\\
&&\hspace{1cm}=\frac12\,\big\langle[(1/2_{(1)}\otimes1/2_{(3)})^{S'}\otimes1/2_{(2)}]^{J'}_{r'}\big|
\sigma_{\pm}^{(1)}\big|
[(1/2_{(1)}\otimes1/2_{(2)})^S\otimes1/2_{(3)}]^{J}_r\big\rangle,
\eea
\item $A^0_{r\to r'},\ A^3_{r\to r'}$
\bea
&&\sum_{s_1,s_2}(1/2,1/2,S';r'-r+s_1,r-s_1-s_2,r'-s_2)(S',1/2,J';r'-s_2,s_2,r')
\nonumber\\&&\hspace{.65cm}(1/2,1/2,S;s_1,s_2,s_1+s_2)
(S,1/2,J;s_1+s_2,r-s_1-s_2,r)\ \delta_{r\,r'}\ \big(\delta_{s_1\,1/2}-
\delta_{s_1\,-1/2}\big)\nonumber\\
&&\hspace{1cm}=\big\langle[(1/2_{(1)}\otimes1/2_{(3)})^{S'}\otimes1/2_{(2)}]^{J'}_{r'}\big|
\sigma_3^{(1)}\big|
[(1/2_{(1)}\otimes1/2_{(2)})^S\otimes1/2_{(3)}]^{J}_r\big\rangle,
\eea
\item $A^1_{r\to r'},\ A^2_{r\to r'}$
\bea
&&\sum_{s_1,s_2}(1/2,1/2,S';r'-r+s_1,r-s_1-s_2,r'-s_2)(S',1/2,J';r'-s_2,s_2,r')
\nonumber\\&&\hspace{.65cm}(1/2,1/2,S;s_1,s_2,s_1+s_2)
(S,1/2,J;s_1+s_2,r-s_1-s_2,r)\ \delta_{r'-r\ \pm1}\nonumber\\
&&\hspace{1cm}\frac12\,\big\langle[(1/2_{(1)}\otimes1/2_{(3)})^{S'}\otimes1/2_{(2)}]^{J'}_{r'}\big|
\sigma_{\pm}^{(1)}\big|
[(1/2_{(1)}\otimes1/2_{(2)})^S\otimes1/2_{(3)}]^{J}_r\big\rangle,
\eea
\end{enumerate}
where $\big|[(1/2_{(1)}\otimes1/2_{(2)})^S\otimes1/2
_{(3)}]^{J}_r\big\rangle$ represents a spin state in which 
quarks 1 and 2 couple
to spin $S$ and then couple with quark 3 to a final state of total spin $J$
and projection $r$. Similarly  $\big|[(1/2_{(1)}\otimes1/2_{(3)})^{S'}\otimes1/2
_{(2)}]^{J'}_{r'}\big\rangle$ is a spin state in which quarks 1 and 3 couple
to spin $S'$ and then couple with quark 2 to a final state of total spin $J'$
and projection $r'$. Besides $\vec\sigma^{(1)}$ is the spin operator for quark 1
being $\sigma^{(1)}_{\pm}=\sigma^{(1)}_1\pm i\sigma^{(1)}_2$. Use of the Wigner-Eckart theorem allows us to immediately obtain
\bea
V^0_{r\to r'}=V^0_{1/2\to1/2}\ \delta_{r\,r'}\ \ &,&
\ \ A^0_{r\to r'}=A^0_{r\to r}\ \delta_{r\,r'},\nonumber\\
V^3_{r\to r'}=V^3_{1/2\to1/2}\ \delta_{r\,r'}\ \ &,&
\ \ A^3_{r\to r'}=A^3_{r\to r}\ \delta_{r\,r'},\nonumber\\
V^1_{r\to r}=0\ \ &,&\ \ A^1_{r\to r}=0,\nonumber\\
V^2_{r\to r}=0\ \ &,&\ \ A^2_{r\to r}=0,
\label{eq:vamer}
\eea
which are valid for all cases under study.
Further relations are quoted in the following.

In terms of matrix elements, the different form factors for the $1/2\to1/2$,
$1/2\to3/2$ and $3/2\to1/2$ can be evaluated as
\begin{enumerate}
\item $\ 1/2\to 1/2$ transitions
\bea
 F_1&=&-\sqrt{\frac{E'+M'}{2M}}\frac1{|\vec q\,|}V^1_{-1/2\to
1/2},\nonumber\\
 F_2&=&\frac1{\sqrt{(E'+M')2M}}\bigg(V^0_{1/2\to1/2}+\frac{E'}{|\vec q\,|}V^3_{1/2\to
1/2}+\frac{M'}{|\vec q\,|}V^1_{-1/2\to1/2}\bigg),\nonumber\\
 F_3&=&-\frac1{\sqrt{(E'+M')2M}}\frac{M'}{|\vec q\,|}\left( V^3_{1/2\to
1/2}-V^1_{-1/2\to
1/2}\right),\\
 G_1&=&\frac1{\sqrt{(E'+M')2M}}A^1_{-1/2\to
1/2},\nonumber\\
 G_2&=&\sqrt{\frac{E'+M'}{2M}}\frac1{|\vec q\,|}\left(A^0_{1/2\to1/2}
-\frac{M'}{|\vec q\,|}A^1_{-1/2\to
1/2}+\frac{E'}{|\vec q\,|}A^3_{1/2\to1/2}\right),\nonumber\\
 G_3&=&-\sqrt{\frac{E'+M'}{2M}}\frac{M'}{|\vec q\,|^2}\left( A^3_{1/2\to
1/2}-A^1_{-1/2\to
1/2}\right).
\eea

\item $\ 1/2\to3/2$ transitions.
\bea
C_3^V&=&\frac{M'}{|\vec q\,|}\frac{1}{\sqrt{2M(E'+M')}}{\sqrt6}\
V^1_{-1/2\to 1/2},\nonumber\\
C_4^V&=&-\frac{M}{M'}\ C_3^V,\nonumber\\
C_5^V&=&C_6^V=0,
\eea
\bea
C_3^A&=&0,\nonumber\\
C_4^A&=&\frac{1}{\sqrt{(E'+M')2M}}\sqrt\frac32\left[
-\frac{M'}{|\vec q\,|}\left(A^0_{1/2\to1/2}+\frac{E'-M}{|\vec q\,|}\,
A^3_{1/2\to 1/2}\right)+
\frac{2(ME'-M'^2)}{|\vec q\,|^2}A^1_{-1/2\to 1/2}\right],
\nonumber\\
C_5^A&=&\frac{M'}{|\vec
q\,|}\frac{1}{\sqrt{(E'+M')2M}}\sqrt\frac32\left[\ \ \frac{ME'-M'^2}{M^2}
\left(A^0_{1/2\to1/2}+\frac{E'-M}{|\vec q\,|}A^3_{1/2\to 1/2}\right)\right.\nonumber \\
&&\hspace{4cm}+\left.
\frac{2M'(2ME'-M^2-M'^2)}{M^2|\vec q\,|}\
A^1_{-1/2\to 1/2}\right],\nonumber\\
C_6^A&=&\frac{M'}{|\vec
q\,|}\frac{1}{\sqrt{(E'+M')2M}}\sqrt\frac32\left(
A^0_{1/2\to1/2}+\frac{E'}{|\vec q\,|}A^3_{1/2\to 1/2}+
\frac{2M'}{|\vec q\,|}\,A^1_{-1/2\to 1/2}\right).
\eea
In the derivation of the above formulas, the following relations found 
among $1/2\to3/2$ matrix elements have 
 been used 
\bea
V^0_{1/2\to1/2}&=&V^3_{1/2\to1/2}=0,\nonumber\\
V^1_{1/2\to-1/2}&=&V^1_{-1/2\to1/2}\ \ ,\ \ 
V^1_{1/2\to3/2}=\sqrt3\ V^1_{-1/2\to1/2},\nonumber\\
A^1_{1/2\to-1/2}&=&-A^1_{-1/2\to1/2}\ \ , \ \
A^1_{1/2\to3/2}=\sqrt3\ A^1_{-1/2\to1/2}.
\eea
\item $\ 3/2\to 1/2$ transitions.
\bea
&&C_3^V=-\frac{M'}{|\vec q\,|}\frac{1}{\sqrt{2M(E'+M')}}{\sqrt6}\
V^1_{-1/2\to 1/2},\nonumber\\
&&C_4^V=-\frac{M'}{M}\ C_3^V,\nonumber\\
&&C_5^V=C_6^V=0,
\eea
\bea
C_3^A&=&0,\nonumber\\
C_4^A&=&\frac{M'^2}{M|\vec q\,|^2}\frac{1}{\sqrt{(E'+M')2M}}\sqrt\frac32
\left[\,\,\big(\,(E'-M)\,A^3_{1/2\to 1/2}+|\vec q\,|\,A^0_{1/2\to1/2}\big)
-2(E'-M)\,A^1_{-1/2\to 1/2}\right],\nonumber\\
C_5^A&=&\frac{1}{|\vec
q\,|^2}\frac{1}{\sqrt{(E'+M')2M}}\sqrt\frac32\
\left( -(E'-M)^2\,A^3_{1/2\to1/2}+|\vec q\,|(M-E')\,A^0_{1/2\to 1/2}\right.\nonumber\\
&&\hspace{4.25cm}\left.-2(2E'M-M'^2-M^2)A^1_{-1/2\to 1/2}\right),\nonumber\\
C_6^A&=&\frac{M'^2}{|\vec
q\,|^2}\frac{1}{\sqrt{(E'+M')2M}}\sqrt\frac32
\left(\,A^3_{1/2\to1/2}-2\,A^1_{-1/2\to 1/2}\right),
\eea
where again we have  made use of the following relations 
observed between $3/2\to1/2$ matrix
elements
\bea
V^0_{1/2\to1/2}&=&V^3_{1/2\to 1/2}=0,\nonumber\\
 V^1_{3/2\to1/2}&=&\sqrt3\,V^1_{-1/2\to
1/2},\nonumber\\
A^1_{3/2\to1/2}&=&-\sqrt3\,A^1_{-1/2\to
1/2}.
\eea
\end{enumerate}

As mentioned we do not use a form factor decomposition for the 
$3/2\to3/2$ transitions but work directly with the matrix elements. For 
$3/2\to3/2$ transitions, and apart from  the relations in Eq.~(\ref{eq:vamer}), 
we further obtain  that 
 \bea
V^1_{-3/2\to -1/2}&=&\frac{\sqrt3}2V^1_{-1/2\to 1/2}=V^1_{1/2\to 3/2}=
-V^1_{3/2\to 1/2}=-\frac{\sqrt3}2V^1_{1/2\to -1/2}=-V^1_{-1/2\to -3/2}
=\frac{-\sqrt2}{\sqrt5}\,{\cal V},\nonumber\\
V^2_{-3/2\to -1/2}&=&\frac{\sqrt3}2V^2_{-1/2\to 1/2}=V^2_{1/2\to 3/2}=V^2_{3/2\to 1/2}=
\frac{\sqrt3}2V^2_{1/2\to -1/2}=V^2_{-1/2\to -3/2}=i\,\frac{\sqrt2}{\sqrt5}
\,{\cal V},
\eea
\bea
A^0_{-3/2\to -3/2}&=&3A^0_{-1/2\to -1/2}=-3A^0_{1/2\to 1/2}=-A^0_{3/2\to 3/2},\nonumber\\
A^3_{-3/2\to -3/2}&=&3A^3_{-1/2\to -1/2}=-3A^3_{1/2\to 1/2}=-A^3_{3/2\to 3/2},\nonumber\\
A^1_{-3/2\to -1/2}&=&\frac{\sqrt3}2A^1_{-1/2\to 1/2}=A^1_{1/2\to 3/2}=
A^1_{3/2\to 1/2}=\frac{\sqrt3}2A^1_{1/2\to -1/2}=A^1_{-1/2\to -3/2}=
\frac{\sqrt2}{\sqrt5}\,{\cal A},\nonumber\\
A^2_{-3/2\to -1/2}&=&\frac{\sqrt3}2A^2_{-1/2\to 1/2}=A^2_{1/2\to 3/2}=-A^2_{3/2\to 1/2}=
-\frac{\sqrt3}2A^2_{1/2\to -1/2}=-A^2_{-1/2\to -3/2}
=i\,\frac{-\sqrt2}{\sqrt5}\,{\cal A},
\eea
where ${\cal V}$ and ${\cal A}$ stand for reduced matrix elements.\\

In every case we just need to evaluate  three different vector and three different axial
matrix elements that we take to be ${ V}^{ 0}_{1/2\to 1/2},\,{ V}^{ 3}_{1/2\to 1/2},
\,{ V}^{ 1}_{-1/2\to 1/2}$ and ${ A}^{ 0}_{1/2\to 1/2},\,{ A}^{ 3}_{1/2\to 1/2},
\,{ A}^{ 1}_{-1/2\to 1/2}$ respectively.
The  vector  matrix elements have the general structure
\bea
{ V}^{ 0}_{1/2\to 1/2}&=&V^{(0)}_{SF}
\sqrt{2M}\sqrt{2E'}\int d^3Q_1\int d^3Q_2\ 
 \left[\tilde\phi^{(B')}(\vec Q_1-\frac{m_c+m_{l'}}{\overline{M'}}\,\vec q,-\vec Q_1-\vec
 Q_2+\frac{m_{l'}}{\overline{M'}}\,\vec q\,)\right]^*\tilde\phi^{(B)}(\vec Q_1,\vec Q_2)\nonumber\\
&&\times\, \sqrt\frac{(E_l(|\vec{Q}_1-\vec{q}\,
|)+m_l)(E_c(|\vec{Q}_1|)+m_c)}{2E_l(|\vec{Q}_1-\vec{q}\, |)2E_c(|\vec{Q}_1|)}
\left(1+\frac{|\vec{Q}_1|^2-|\vec{q}\, |Q_1^z}{(E_l(|\vec{Q}_1-\vec{q}\, |)+m_l)(E_c(|\vec{Q}_1|)+m_c)}
\right),
\eea
\bea
{ V}^{ 3}_{1/2\to 1/2}&=&V^{(3)}_{SF}
\sqrt{2M}\sqrt{2E'}\int d^3Q_1\int d^3Q_2\ 
 \left[\tilde\phi^{(
 B')}(\vec Q_1-\frac{m_c+m_{l'}}{\overline{M'}}\,\vec q,-\vec Q_1-\vec
 Q_2+\frac{m_{l'}}{\overline{M'}}\,\vec q\,)\right]^*\tilde\phi^{(B)}(\vec Q_1,\vec Q_2)\nonumber\\
&&\times\, \sqrt\frac{(E_l(|\vec{Q}_1-\vec{q}\,
|)+m_l)(E_c(|\vec{Q}_1|)+m_c)}{2E_l(|\vec{Q}_1-\vec{q}\, |)2E_c(|\vec{Q}_1|)}
\left(\frac{Q_1^z}{E_c(|\vec{Q}_1|)+m_c}+\frac{Q_1^z-|\vec{q}\,|}
{E_l(|\vec{Q}_1-\vec{q}\, |)+m_l}
\right),
\eea
\bea
{ V}^{ 1}_{-1/2\to 1/2}&=&V^{(1)}_{SF}
\sqrt{2M}\sqrt{2E'}\int d^3Q_1\int d^3Q_2\ 
 \left[\tilde\phi^{(
 B')}(\vec Q_1-\frac{m_c+m_{l'}}{\overline{M'}}\,\vec q,-\vec Q_1-\vec
 Q_2+\frac{m_{l'}}{\overline{M'}}\,\vec q\,)\right]^*\tilde\phi^{(B)}(\vec Q_1,\vec Q_2)\nonumber\\
&&\times\, \sqrt\frac{(E_l(|\vec{Q}_1-\vec{q}\, |)+m_l)
(E_c(|\vec{Q}_1|)+m_c)}{2E_l(|\vec{Q}_1-\vec{q}\, |)2E_c(|\vec{Q}_1|)}
\nonumber\\
&&\times\, \frac{|\vec{q}\,|(E_c(|\vec{Q}_1|)+m_c)-[E_c(|\vec{Q}_1|)+m_c-
E_l(|\vec{Q}_1-\vec{q}\, |)-m_l]\,Q_1^z}{(E_l(|\vec{Q}_1-\vec{q}\, |)+m_l)
(E_c(|\vec{Q}_1|)+m_c)}.
\eea
The
$V^{(j)}_{SF}$ depend on the flavor and spin structure of the baryons
involved. Their values for the different transitions appear in
Table~\ref{tab:SVAFctosd}. 
\begin{table}
\begin{tabular}{lcccccc}\hline
&\ \ $V^{(0)}_{SF}$\ \ &\ \ $V^{(3)}_{SF}$\ \ &\ \ $V^{(1)}_{SF}$\ \ &\ \ $A^{(0)}_{SF}$\ \ &\ \ $A^{(3)}_{SF}$\ \ &\ \ $A^{(1)}_{SF}$\ \ 
\vspace{.1cm}\\\hline\\
$\Xi_{cb}^{+}\to\Xi^0_b$
 &$\frac{\sqrt3}{2}$&$\frac{\sqrt3}{2}$&$\frac{-1}{2\sqrt3}$
 &$\frac{1}{2\sqrt3}$&$\frac{1}{2\sqrt3}$&$\frac{1}{2\sqrt3}$\\
 
$\Xi_{cb}^{0}\to\Xi^-_b$
 &$\frac{\sqrt3}{2}$&$\frac{\sqrt3}{2}$&$\frac{-1}{2\sqrt3}$
 &$\frac{1}{2\sqrt3}$&$\frac{1}{2\sqrt3}$&$\frac{1}{2\sqrt3}$\\

$\Xi_{cb}^{+}\to\Xi^{\prime\,0}_b$
 &$\frac{1}{2}$&$\frac{1}{2}$&$\frac{-5}{6}$
 &$\frac{5}{6}$&$\frac{5}{6}$&$\frac{5}{6}$\\

$\Xi_{cb}^{0}\to\Xi^{\prime\,-}_b$
 &$\frac{1}{2}$&$\frac{1}{2}$&$\frac{-5}{6}$
 &$\frac{5}{6}$&$\frac{5}{6}$&$\frac{5}{6}$\\

$\Xi_{cb}^{+}\to\Xi^{*\,0}_b$
 &$0$&$0$&$\frac{-1}{3\sqrt2}$
 &$\frac{-\sqrt2}{3}$&$\frac{-\sqrt2}{3}$&$\frac{1}{3\sqrt2}$\\

$\Xi_{cb}^{0}\to\Xi^{*\,-}_b$
 &$0$&$0$&$\frac{-1}{3\sqrt2}$
 &$\frac{-\sqrt2}{3}$&$\frac{-\sqrt2}{3}$&$\frac{1}{3\sqrt2}$\\

$\Xi_{cb}^{\prime\,+}\to\Xi^0_b$
 &$\frac{1}{2}$&$\frac{1}{2}$&$\frac{1}{2}$
 &$\frac{-1}{2}$&$\frac{-1}{2}$&$\frac{-1}{2}$\\
 
$\Xi_{cb}^{\prime\,0}\to\Xi^-_b$
 &$\frac{1}{2}$&$\frac{1}{2}$&$\frac{1}{2}$
 &$\frac{-1}{2}$&$\frac{-1}{2}$&$\frac{-1}{2}$\\

$\Xi_{cb}^{\prime\,+}\to\Xi^{\prime\,0}_b$
 &$\frac{-\sqrt3}{2}$&$\frac{-\sqrt3}{2}$&$\frac{1}{2\sqrt3}$
 &$\frac{-1}{2\sqrt3}$&$\frac{-1}{2\sqrt3}$&$\frac{-1}{2\sqrt3}$\\
 
$\Xi_{cb}^{\prime\,0}\to\Xi^{\prime\,-}_b$
 &$\frac{-\sqrt3}{2}$&$\frac{-\sqrt3}{2}$&$\frac{1}{2\sqrt3}$
 &$\frac{-1}{2\sqrt3}$&$\frac{-1}{2\sqrt3}$&$\frac{-1}{2\sqrt3}$\\
 
$\Xi_{cb}^{\prime\,+}\to\Xi^{*\,0}_b$
 &$0$&$0$&$\frac{-1}{\sqrt6}$
 &$\frac{-\sqrt2}{\sqrt3}$&$\frac{-\sqrt2}{\sqrt3}$&$\frac{1}{\sqrt6}$\\

$\Xi_{cb}^{\prime\,0}\to\Xi^{*\,-}_b$
 &$0$&$0$&$\frac{-1}{\sqrt6}$
 &$\frac{-\sqrt2}{\sqrt3}$&$\frac{-\sqrt2}{\sqrt3}$&$\frac{1}{\sqrt6}$\\

$\Xi_{cb}^{*\,+}\to\Xi^{0}_b$
 &$0$&$0$&$\frac{-1}{\sqrt6}$
 &$\frac{\sqrt2}{\sqrt3}$&$\frac{\sqrt2}{\sqrt3}$&$\frac{1}{\sqrt6}$\\

$\Xi_{cb}^{*\,0}\to\Xi^{-}_b$
 &$0$&$0$&$\frac{-1}{\sqrt6}$
 &$\frac{\sqrt2}{\sqrt3}$&$\frac{\sqrt2}{\sqrt3}$&$\frac{1}{\sqrt6}$\\

$\Xi_{cb}^{*\,+}\to\Xi^{\prime\,0}_b$
 &$0$&$0$&$\frac{1}{3\sqrt2}$
 &$\frac{-\sqrt2}{3}$&$\frac{-\sqrt2}{3}$&$\frac{-1}{3\sqrt2}$\\

$\Xi_{cb}^{*\,0}\to\Xi^{\prime\,-}_b$
 &$0$&$0$&$\frac{1}{3\sqrt2}$
 &$\frac{-\sqrt2}{3}$&$\frac{-\sqrt2}{3}$&$\frac{-1}{3\sqrt2}$\\

$\Xi^{*\,+}_{cb}\to\Xi^{*\,0}_b$
 &$-1$&$-1$&$\frac{2}{3}$
 &$\frac{-1}{3}$&$\frac{-1}{3}$&$\frac{-2}{3}$\\

$\Xi^{*\,0}_{cb}\to\Xi^{*\,-}_b$
 &$-1$&$-1$&$\frac{2}{3}$
 &$\frac{-1}{3}$&$\frac{-1}{3}$&$\frac{-2}{3}$\\

$\Omega^{0}_{cb}\to\Omega^{-}_b$
 &$\frac1{\sqrt2}$&$\frac1{\sqrt2}$&$\frac{-5}{3\sqrt2}$
 &$\frac{5}{3\sqrt2}$&$\frac{5}{3\sqrt2}$&$\frac{5}{3\sqrt2}$\\

$\Omega^{0}_{cb}\to\Omega^{*\,-}_b$
 &$0$&$0$&$\frac{-1}{3}$
 &$\frac{-2}{3}$&$\frac{-2}{3}$&$\frac{1}{3}$\\

$\Omega^{\prime\,0}_{cb}\to\Omega^{-}_b$
 &$\frac{-\sqrt3}{\sqrt2}$&$\frac{-\sqrt3}{\sqrt2}$&$\frac{1}{\sqrt6}$
 &$\frac{-1}{\sqrt6}$&$\frac{-1}{\sqrt6}$&$\frac{-1}{\sqrt6}$\\

$\Omega^{\prime\,0}_{cb}\to\Omega^{*\,-}_b$
 &$0$&$0$&$\frac{-1}{\sqrt3}$
 &$\frac{-2}{\sqrt3}$&$\frac{-2}{\sqrt3}$&$\frac{1}{\sqrt3}$\\
 
$\Omega^{*\,0}_{cb}\to\Omega^{-}_b$
 &$0$&$0$&$\frac{1}{3}$
 &$\frac{-2}{3}$&$\frac{-2}{3}$&$\frac{-1}{3}$\\

$\Omega^{*\,0}_{cb}\to\Omega^{*\,-}_b$
 &${-\sqrt2}$&${-\sqrt2}$&$\frac{2\sqrt2}{3}$
 &$\frac{-\sqrt2}{3}$&$\frac{-\sqrt2}{3}$&$\frac{-2\sqrt2}{3}$\\
\\
 
\hline
\end{tabular}
%
%
%
%
\hspace{.5cm}
\begin{tabular}{lcccccc}\hline
&\ \ $V^{(0)}_{SF}$\ \ &\ \ $V^{(3)}_{SF}$\ \ &\ \ $V^{(1)}_{SF}$\ \ &\ \ $A^{(0)}_{SF}$\ \ &\ \ $A^{(3)}_{SF}$\ \ &\ \ $A^{(1)}_{SF}$\ \ 
\vspace{.1cm}\\\hline\\
$\Xi_{cb}^{+}\to\Lambda^0_b$
 &$\frac{\sqrt3}{2}$&$\frac{\sqrt3}{2}$&$\frac{-1}{2\sqrt3}$
 &$\frac{1}{2\sqrt3}$&$\frac{1}{2\sqrt3}$&$\frac{1}{2\sqrt3}$\\
 
$\Xi_{cb}^{+}\to\Sigma^0_b$
 &$\frac{1}{2}$&$\frac{1}{2}$&$\frac{-5}{6}$
 &$\frac{5}{6}$&$\frac{5}{6}$&$\frac{5}{6}$\\

$\Xi_{cb}^{0}\to\Sigma^-_b$
 &$\frac{1}{\sqrt2}$&$\frac{1}{\sqrt2}$&$\frac{-5}{3\sqrt2}$
 &$\frac{5}{3\sqrt2}$&$\frac{5}{3\sqrt2}$&$\frac{5}{3\sqrt2}$\\

$\Xi_{cb}^{+}\to\Sigma^{*\,0}_b$
 &$0$&$0$&$\frac{-1}{3\sqrt2}$
 &$\frac{-\sqrt2}{3}$&$\frac{-\sqrt2}{3}$&$\frac{1}{3\sqrt2}$\\

$\Xi_{cb}^{0}\to\Sigma^{*\,-}_b$
 &$0$&$0$&$\frac{-1}{3}$
 &$\frac{-2}{3}$&$\frac{-2}{3}$&$\frac{1}{3}$\\

$\Xi^{\prime\,+}_{cb}\to\Lambda^0_b$
 &$\frac{1}{2}$&$\frac{1}{2}$&$\frac{1}{2}$
 &$\frac{-1}{2}$&$\frac{-1}{2}$&$\frac{-1}{2}$\\
 
$\Xi^{\prime\,+}_{cb}\to\Sigma^0_b$
 &$\frac{-\sqrt3}{2}$&$\frac{-\sqrt3}{2}$&$\frac{1}{2\sqrt3}$
 &$\frac{-1}{2\sqrt3}$&$\frac{-1}{2\sqrt3}$&$\frac{-1}{2\sqrt3}$\\

$\Xi^{\prime\,0}_{cb}\to\Sigma^-_b$
 &$\frac{-\sqrt3}{\sqrt2}$&$\frac{-\sqrt3}{\sqrt2}$&$\frac{1}{\sqrt6}$
 &$\frac{-1}{\sqrt6}$&$\frac{-1}{\sqrt6}$&$\frac{-1}{\sqrt6}$\\

$\Xi^{\prime\,+}_{cb}\to\Sigma^{*\,0}_b$
 &$0$&$0$&$\frac{-1}{\sqrt6}$
 &$\frac{-\sqrt2}{\sqrt3}$&$\frac{-\sqrt2}{\sqrt3}$&$\frac{1}{\sqrt6}$\\

$\Xi^{\prime\,0}_{cb}\to\Sigma^{*\,-}_b$
 &$0$&$0$&$\frac{-1}{\sqrt3}$
 &$\frac{-2}{\sqrt3}$&$\frac{-2}{\sqrt3}$&$\frac{1}{\sqrt3}$\\
 
$\Xi^{*\,+}_{cb}\to\Lambda^{0}_b$
 &$0$&$0$&$\frac{-1}{\sqrt6}$
 &$\frac{\sqrt2}{\sqrt3}$&$\frac{\sqrt2}{\sqrt3}$&$\frac{1}{\sqrt6}$\\
 
$\Xi^{*\,+}_{cb}\to\Sigma^{0}_b$
 &$0$&$0$&$\frac{1}{3\sqrt2}$
 &$\frac{-\sqrt2}{3}$&$\frac{-\sqrt2}{3}$&$\frac{-1}{3\sqrt2}$\\
 
$\Xi^{*\,0}_{cb}\to\Sigma^{-}_b$
 &$0$&$0$&$\frac{1}{3}$
 &$\frac{-2}{3}$&$\frac{-2}{3}$&$\frac{-1}{3}$\\
 
$\Xi^{*\,+}_{cb}\to\Sigma^{*\,0}_b$
 &$-1$&$-1$&$\frac{2}{3}$
 &$\frac{-1}{3}$&$\frac{-1}{3}$&$\frac{-2}{3}$\\

$\Xi^{*\,0}_{cb}\to\Sigma^{*\,-}_b$
 &$-\sqrt2$&$-\sqrt2$&$\frac{2\sqrt2}{3}$
 &$\frac{-\sqrt2}{3}$&$\frac{-\sqrt2}{3}$&$\frac{-2\sqrt2}{3}$\\

$\Omega^{0}_{cb}\to\Xi^{-}_b$
 &$\frac{-\sqrt3}2$&$\frac{-\sqrt3}2$&$\frac{1}{2\sqrt3}$
 &$\frac{-1}{2\sqrt3}$&$\frac{-1}{2\sqrt3}$&$\frac{-1}{2\sqrt3}$\\

$\Omega^{0}_{cb}\to\Xi^{\prime\,-}_b$
 &$\frac{1}2$&$\frac{1}2$&$\frac{-5}{6}$
 &$\frac{5}{6}$&$\frac{5}{6}$&$\frac{5}{6}$\\

$\Omega^{0}_{cb}\to\Xi^{*\,-}_b$
 &$0$&$0$&$\frac{-1}{3\sqrt2}$
 &$\frac{-2}{3\sqrt2}$&$\frac{-2}{3\sqrt2}$&$\frac{1}{3\sqrt2}$\\

$\Omega^{\prime\,0}_{cb}\to\Xi^{-}_b$
 &$\frac{-1}2$&$\frac{-1}2$&$\frac{-1}{2}$
 &$\frac{1}{2}$&$\frac{1}{2}$&$\frac{1}{2}$\\

$\Omega^{\prime\,0}_{cb}\to\Xi^{\prime\,-}_b$
 &$\frac{-\sqrt3}2$&$\frac{-\sqrt3}2$&$\frac{1}{2\sqrt3}$
 &$\frac{-1}{2\sqrt3}$&$\frac{-1}{2\sqrt3}$&$\frac{-1}{2\sqrt3}$\\

$\Omega^{\prime\,0}_{cb}\to\Xi^{*\,-}_b$
 &$0$&$0$&$\frac{-1}{\sqrt6}$
 &$\frac{-\sqrt2}{\sqrt3}$&$\frac{-\sqrt2}{\sqrt3}$&$\frac{1}{\sqrt6}$\\

$\Omega^{*\,0}_{cb}\to\Xi^{-}_b$
 &$0$&$0$&$\frac{1}{\sqrt6}$
 &$\frac{-\sqrt2}{\sqrt3}$&$\frac{-\sqrt2}{\sqrt3}$&$\frac{-1}{\sqrt6}$\\

$\Omega^{*\,0}_{cb}\to\Xi^{\prime\,-}_b$
 &$0$&$0$&$\frac{1}{3\sqrt2}$
 &$\frac{-\sqrt2}{3}$&$\frac{-\sqrt2}{3}$&$\frac{-1}{3\sqrt2}$\\

$\Omega^{*\,0}_{cb}\to\Xi^{*\,-}_b$
 &$-1$&$-1$&$\frac{2}{3}$
 &$\frac{-1}{3}$&$\frac{-1}{3}$&$\frac{-2}{3}$\\\\

\hline
\end{tabular}
\caption{$V^{(j)}_{SF}$ and $A^{(j)}_{SF}$ spin-flavor factors
for  for $c\to s$ (left panel) and $c\to d$ (right panel) transitions.}
\label{tab:SVAFctosd}
\end{table}
Similarly, for the axial matrix elements we have
\bea
{ A}^{ 0}_{1/2\to 1/2}&=&A^{(0)}_{SF}
\sqrt{2M}\sqrt{2E'}\int d^3Q_1\int d^3Q_2\ 
 \left[\tilde\phi^{(
 B')}(\vec Q_1-\frac{m_c+m_{l'}}{\overline{M'}}\,\vec q,-\vec Q_1-\vec
 Q_2+\frac{m_{l'}}{\overline{M'}}\,\vec q\,)\right]^*\tilde\phi^{(B)}(\vec Q_1,\vec Q_2)\nonumber\\
&&\times\,  \sqrt\frac{(E_l(|\vec{Q}_1-\vec{q}\, |)+m_l)
(E_c(|\vec{Q}_1|)+m_c)}{2E_l(|\vec{Q}_1-\vec{q}\, |)2E_c(|\vec{Q}_1|)}
\left(\frac{{Q}_1^z}{E_c(|\vec{Q}_1|)+m_c}
+
\frac{{Q}_1^z-|\vec{q}\, |}{E_l(|\vec{Q}_1-\vec{q}\, |)+m_l}
\right),
\eea
\bea
{ A}^{ 3}_{1/2\to 1/2}&=&A^{(3)}_{SF}
\sqrt{2M}\sqrt{2E'}\int d^3Q_1\int d^3Q_2\ 
 \left[\tilde\phi^{(
 B')}(\vec Q_1-\frac{m_c+m_{l'}}{\overline{M'}}\,\vec q,-\vec Q_1-\vec
 Q_2+\frac{m_{l'}}{\overline{M'}}\,\vec q\,)\right]^*\tilde\phi^{(B)}(\vec Q_1,\vec Q_2)\nonumber\\
&&\times\,  \sqrt\frac{(E_l(|\vec{Q}_1-\vec{q}\, |)+m_l)
(E_c(|\vec{Q}_1|)+m_c)}{2E_n(|\vec{Q}_1-\vec{q}\, |)2E_c(|\vec{Q}_1|)}
\left(1-\frac{|\vec{Q}_1|^2-|\vec{q}\, |Q_1^z-2Q_1^z(Q_1^z-|\vec q\,|)}
{(E_l(|\vec{Q}_1-\vec{q}\, |)+m_l)
(E_c(|\vec{Q}_1|)+m_c)}
\right),
\eea
\bea
{ A}^{ 1}_{-1/2\to 1/2}&=&A^{(1)}_{SF}
\sqrt{2M}\sqrt{2E'}\int d^3Q_1\int d^3Q_2\ 
 \left[\tilde\phi^{(
 B')}(\vec Q_1-\frac{m_c+m_{l'}}{\overline{M'}}\,\vec q,-\vec Q_1-\vec
 Q_2+\frac{m_{l'}}{\overline{M'}}\,\vec q\,)\right]^*\tilde\phi^{(B)}(\vec Q_1,\vec Q_2)\nonumber\\
&&\times\,  \sqrt\frac{(E_l(|\vec{Q}_1-\vec{q}\, |)+m_l)
(E_c(|\vec{Q}_1|)+m_c)}{2E_l(|\vec{Q}_1-\vec{q}\, |)2E_c(|\vec{Q}_1|)}
\ \left(1-\frac{|\vec{Q}_1|^2-|\vec{q}\, |Q_1^z-2Q_1^x(Q_1^x-iQ_1^y)}
{(E_l(|\vec{Q}_1-\vec{q}\, |)+m_l)
(E_c(|\vec{Q}_1|)+m_c)}
\right),
\eea
where the $A^{(j)}_{SF}$ axial spin-flavor factors can be found in 
Table~\ref{tab:SVAFctosd}. Note that due to the symmetry properties already 
discussed, 
the integral in
$2Q_1^xQ_1^x$ in ${ A}^{ 1}_{-1/2\to 1/2}$ es equivalent to an integral in
$|\vec Q_1|^2-(Q_1^z)^2$, while the integral
in $2Q_1^xQ_1^y$ is identically zero. 

As already said, when the final baryon has just one $s$ quark
then the $\tilde\phi^{( B')}$ above should be interpreted as $\tilde\phi^{(
B')}_{sn}$ or $\tilde\phi^{( B')}_{ds}$,  for the case of 
 $c\to s$ or $c\to d$ transitions, respectively.

\end{document}